\documentclass[twocolumn,prx,superscriptaddress,10pt]{revtex4-1}

\usepackage[usenames]{color}
\usepackage{graphicx}
\usepackage{amsmath}
\usepackage{amssymb}
\usepackage[normalem]{ulem}
\usepackage{mathrsfs}
\usepackage{bm}
\usepackage{epstopdf}
\usepackage{xcolor}
\usepackage{soul}
\usepackage{bbold}
\usepackage{mathrsfs}
\usepackage[section]{placeins}
\usepackage[multiple]{footmisc}
\usepackage{caption}

\usepackage{color}
\definecolor{red}{rgb}{0,0,0}
\definecolor{blue}{rgb}{
0,0,0.75}
\definecolor{green}{rgb}{0,0.5,0}

\newcommand{\red}[1]{{\color{red} #1}}

\setstcolor{black}

\usepackage{verbatim}

\begin{document}


\title{Irreversibility and symmetry breaking in the creation and annihilation of defects in active living matter}

\author{A. Be’er}
\affiliation{Zuckerberg Institute for Water Research, The Jacob Blaustein Institutes for Desert Research, 
Ben-Gurion University of the Negev, Sede Boqer Campus, Midreshet Ben‑Gurion, Israel}
\affiliation{The Department of Physics, Ben-Gurion University of the Negev, Beer‑Sheva, Israel}

\author{E.D. Neimand}
\affiliation{The Swiss Institute for Dryland Environmental and Energy Research, The Jacob Blaustein Institutes for Desert Research, Ben-Gurion University of the Negev, Sede Boqer Campus, Midreshet Ben-Gurion, Israel}

\author{\red{Y. Agarwal}}
\affiliation{Department of Theoretical Physics, University of Geneva, Geneva, Switzerland}

\author{D. Corbett}
\affiliation{Department of Theoretical Physics, University of Geneva, Geneva, Switzerland}

\author{D.J.G. Pearce}
\affiliation{Department of Theoretical Physics, University of Geneva, Geneva, Switzerland}

\author{G. Ariel}
\affiliation{Department of Mathematics, Bar-Ilan University, Ramat‑Gan, Israel}

\author{V.~Yashunsky}
\affiliation{The Swiss Institute for Dryland Environmental and Energy Research, The Jacob Blaustein Institutes for Desert Research, Ben-Gurion University of the Negev, Sede Boqer Campus, Midreshet Ben-Gurion, Israel}
\email{yashunsk@bgu.ac.il}

\date{\today}

\begin{abstract}
Active living matter continuously creates and annihilates topological defects in a process that remains poorly understood. Here, we investigate these dynamics in two distinct active living systems—swarming bacteria and human bronchial epithelial cells. Despite their entirely different evolutionary origins, biological functions, and physical scales, both systems exhibit half-integer defects, consistent with the nematic phase. 
However, in contrast to active nematic theory, we find that defect creation and annihilation undergoes spatial symmetry breaking.
We propose that this results \red{stems} from a fundamental dualism between nematic structural organization and generated polar forces, which are intrinsic to living systems. Furthermore,  estimation of entropy production reveals that creation and annihilation are not reversed processes.
Our findings challenge conventional nematic models and emphasize the role of defect-mediated dynamics in non-equilibrium biological systems as a major source of entropy production.
\end{abstract}

\maketitle


\section{Introduction}

\noindent
Active nematic materials are typically comprised of elongated components that generate stress at the microscopic scale, leading to large-scale motion \cite{sanchez2012spontaneous, guillamat2018active, tan2019topological, kumar2018tunable}. 
Unlike passive systems, active matter remains in a state of perpetual motion and reorganization, which is driven by internally generated forces \cite{sokolov2010swimming, beta2023actin}. A hallmark of active nematic materials is the continuous creation and annihilation of topological defects \cite{marchetti2013hydrodynamics}, which result from a balance between active and elastic forces \cite{giomi2015geometry, zhou2014living, serra2023defect, guillamat2017taming}.
Of particular interest are living systems, such as bacterial collectives \cite{li2019data, yashunsky2024topological, shimaya2022tilt, copenhagen2021topological, genkin2017topological} and tissue cells \cite{saw2017topological, balasubramaniam2021investigating, blanch2018turbulent}, because their physical properties are also related to their biological functions \cite{be2019statistical, aranson2022bacterial, doostmohammadi2022physics}. In living matter, active stress originates from metabolism and the consumption of chemical fuel, which enables bacteria and cells to move and generate forces on their neighbors. Elasticity arises from an elongated cellular shape, leading to liquid-crystal-like properties~\cite{be2019statistical, aranson2022bacterial, doostmohammadi2022physics}. The resulting defects are meso-scopic objects involving tens to hundreds of cells and governed by intrinsic spatiotemporal scales that are independent of system size \cite{li2019data, yashunsky2024topological, shimaya2022tilt, copenhagen2021topological, genkin2017topological, saw2017topological, balasubramaniam2021investigating, blanch2018turbulent}.

Experiments with bacterial colonies and cell monolayers consistently reveal half-integer defects, in accordance with nematic symmetry \cite{li2019data, yashunsky2024topological, shimaya2022tilt, copenhagen2021topological, genkin2017topological, duclos2017topological, blanch2018turbulent, saw2017topological, kawaguchi2017topological, balasubramaniam2021investigating, ienaga2023geometric}. \red{Recent studies suggested that nematic symmetry can also manifest in systems with clear polar directionality \cite{venkatesh2025interplay}, such as self-propelled rod models \cite{shi2018self, grossmann2020particle,de2025self} and rod-like bacteria gliding on surfaces \cite{han2025local}. Under certain conditions, active nematic systems can even acquire polar features, indicating that nematic and polar symmetries can coexist \cite{huber2018emergence, meacock2021bacteria, maroudas2021topological, amiri2022unifying, lacroix2024emergence,vafa2025phase}. 
In \cite{han2025local}, polar stresses in the nematically organized {\it Myxococcus xanthus} colonies were shown to be associated with multi-layer formation.  
Understanding how living systems organize may depend on this interplay of symmetries, which could underlie processes such as biofilm formation and tissue organization, yet it remains largely untested experimentally.} 

Defect interactions have been primarily investigated through numerical simulations~\cite{thampi2014vorticity, shankar2018defect, patelli2019understanding, bonn2022fluctuation, pearce2021orientational}. Elastic forces between defects result in polar-like interactions, as well as additional forces and torques that rely on their relative orientation \cite{shankar2022topological,pearce2021properties,tang2017orientation,vromans2016orientational}. It has been observed that nematic active stresses also lead to the generation of hydrodynamic flows around the defects which results in self-propulsion of $+^1/_2$ defects. Rotational diffusion can reduce persistence in $+^1/_2$ defect motion, preventing unbinding from $-^1/_2$ counterparts, and affecting defect density in steady states \cite{shankar2018defect}. However, much of the theoretical framework remains experimentally unverified. 

Recent experiments on microtubule-based active nematics and swarming bacteria have uncovered a broad spectrum of defect organization, ranging from giant number fluctuations to hyperuniformity \cite{yashunsky2024topological, de2025hidden, pearce2021orientational}. In fibrosarcoma cell cultures, $+^1/_2$ defects align toward the edge of the colony, generating \red{chiral} edge flows through local chiral active stress \cite{yashunsky2022chiral}. \red{Spontaneous mirror (chiral) symmetry breaking has been demonstrated for $+^1/_2$ defects, which specifically concentrate at the boundary between vorticity-dominated and strain-rate-dominated regions in microtubule-kinesin films, highlighting their strong coupling to large-scale hydrodynamic fluxes within the material \cite{head2024spontaneous}.} However, the creation and annihilation processes, which we identify below as primary sources of energy injection and dissipation \red{in living systems, are underexplored, necessitating experimental validation of current theoretical predictions.} 

In this paper, we experimentally study swarming bacteria \cite{be2019statistical} and human bronchial epithelial cells (HBECs) \cite{blanch2018turbulent}—two distinct systems that are known to display nematic half-integer defects. Our findings reveal that the creation and annihilation of defects follow distinct, previously unrecognized trajectories that \red{spontaneously break mirror symmetry}—a feature absent in standard active nematic models. We also find that these trajectories exhibit time-reversal symmetry breaking, a hallmark of non-equilibrium processes. Statistical analysis reveals that defect creation and annihilation are \red{a major source of} entropy production in the bulk, underscoring their irreversible nature and critical role in active matter organization. \red{Finally, we propose a theoretical model that explains these observations taking into account directional (polar) self-propulsion in bacteria and cells}.  

\section{Results}
\begin{figure}[h] 
	\centering
	\includegraphics[width=.5\textwidth]{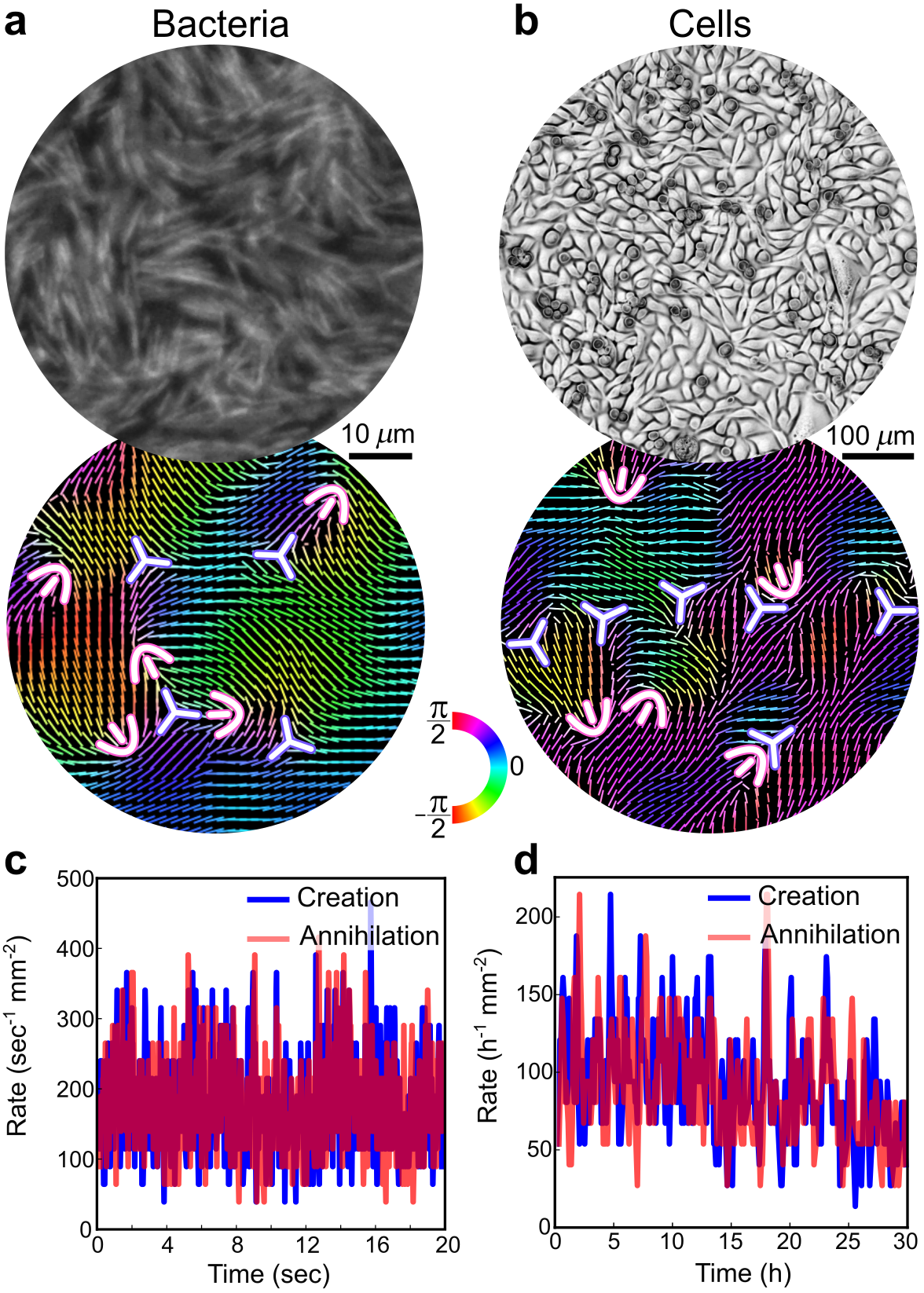}
        \captionsetup{justification=raggedright} 
	\caption{\textbf{Director field and nematic defects in swarming \textit{Bacillus subtilis} and human bronchial epithelial cells (HBECs) monolayers.}
    The nematic director field, colored according to the orientations (ranging from $-\pi/2$ to $\pi/2$), and half-integer defects in (a) swarming \textit{B. subtilis} and (b) HBECs. $+^1/_2$ types are represented as pink comet shapes and $-^1/_2$ as violet tripods. (c, d) Rates of pair-creation (blue) and annihilation (red) of oppositely charged $\pm^1/_2$ defects; (c) swarming \textit{B. subtilis} and (d) HBECs.}
	\label{fig:F1} 
\end{figure}

\subsection*{Half-integer defects in bacteria and cells}
We study the creation and annihilation of defects in two highly active, quasi-turbulent living systems of swarming \textit{Bacillus subtilis} bacteria \cite{yashunsky2024topological} and human bronchial epithelial cells (HBECs) \cite{blanch2018turbulent}. Despite the extreme disparity in the time scales and dimension 
(Fig. \ref{fig:F1}), both systems share fundamental characteristics of active extensile nematics \cite{giomi2015geometry, martinez2021scaling}: (i) A steady-state mixture of half-integer defects (Figs. \ref{fig:F1}a-b), which is obtained through continuing creation and annihilation events (Figs. \ref{fig:F1}c-d), (ii) a counter-rotating vortex pattern around defects with  (iii) an exponential distribution of vortex sizes \cite{yashunsky2022chiral, blanch2018turbulent}. 

\begin{figure*}[ht!] 
	\centering
	\includegraphics[width=.75\textwidth]{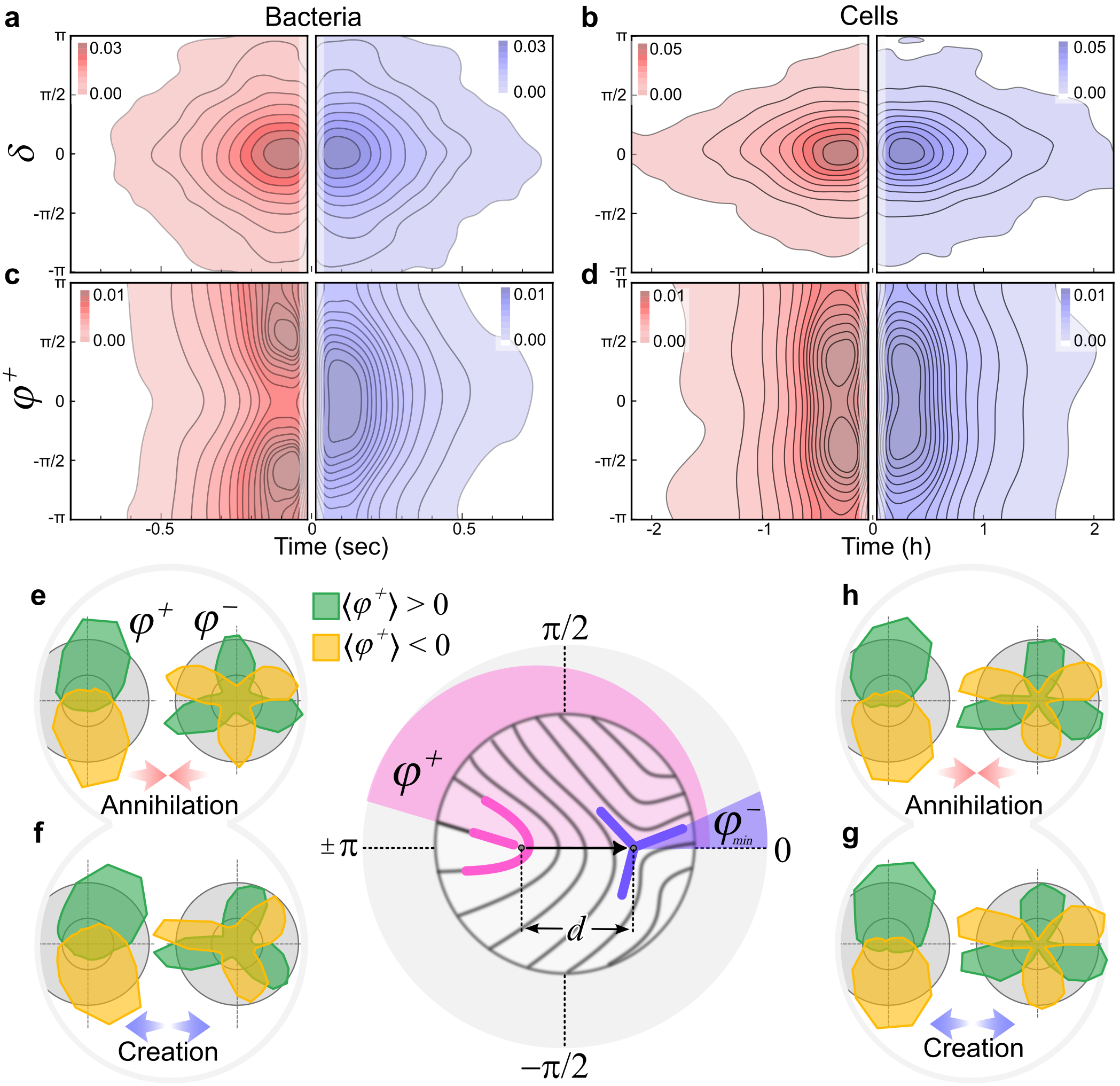} 
        \captionsetup{justification=raggedright} 
	\caption{\textbf{Configuration of $+^1/_2$ and $-^1/_2$ defect-pairs during creation and annihilation.}
    Center sketch: Definitions of orientational angles $\varphi^+$ \red{(pink)} and $\varphi^-$ \red{(violet)} for paired $+^1/_2$ and $-^1/_2$ defects. \red{These angles are defined relative to the vector pointing from the $+^1/_2$ to the $-^1/_2$ defect, shown by the solid black arrow.} \textbf{(a, b)} Distribution of the pair orientation offset angle, $\delta=\varphi^+ -3\varphi^- -\pi$, as a function of time for creation (blue) and annihilation (red) events. $\delta=0$ implies that the nematic director fields of the paired-defects are in phase. \textbf{(c, d)} Distribution of $\varphi^+$ as a function of time for the creation (blue) and annihilation (red) events. The creation/annihilation instant is indicated by $t=0$, with negative time indicating time to annihilation, and positive time is from creation. Red and blue color intensities indicate probability density. \red{The bright line close to $t=0$ indicates a region in which the distance between the defects is up to half cell-body length ($3-4~\mu m$ for bacteria and $7-8 ~\mu m$ for cells). At these distances the continuous nematic field is not well defined and the discrete nature of particles is of concern.} \textbf{(e-g)} \red{Angular distribution of $\varphi^+$ and $\varphi^-$ for defect pairs at small separation ($d<15 ~\mu m$ for bacteria and $d<35 ~\mu m$ for cells). Pairs are classified as ``up'' ($\langle \varphi^+ \rangle >0$, green) or ``down'' ($\langle \varphi^+ \rangle <0$, yellow).}
    }
	\label{fig:F2} 
\end{figure*}

\begin{figure*}
	\centering
	\includegraphics[width=.75\textwidth]{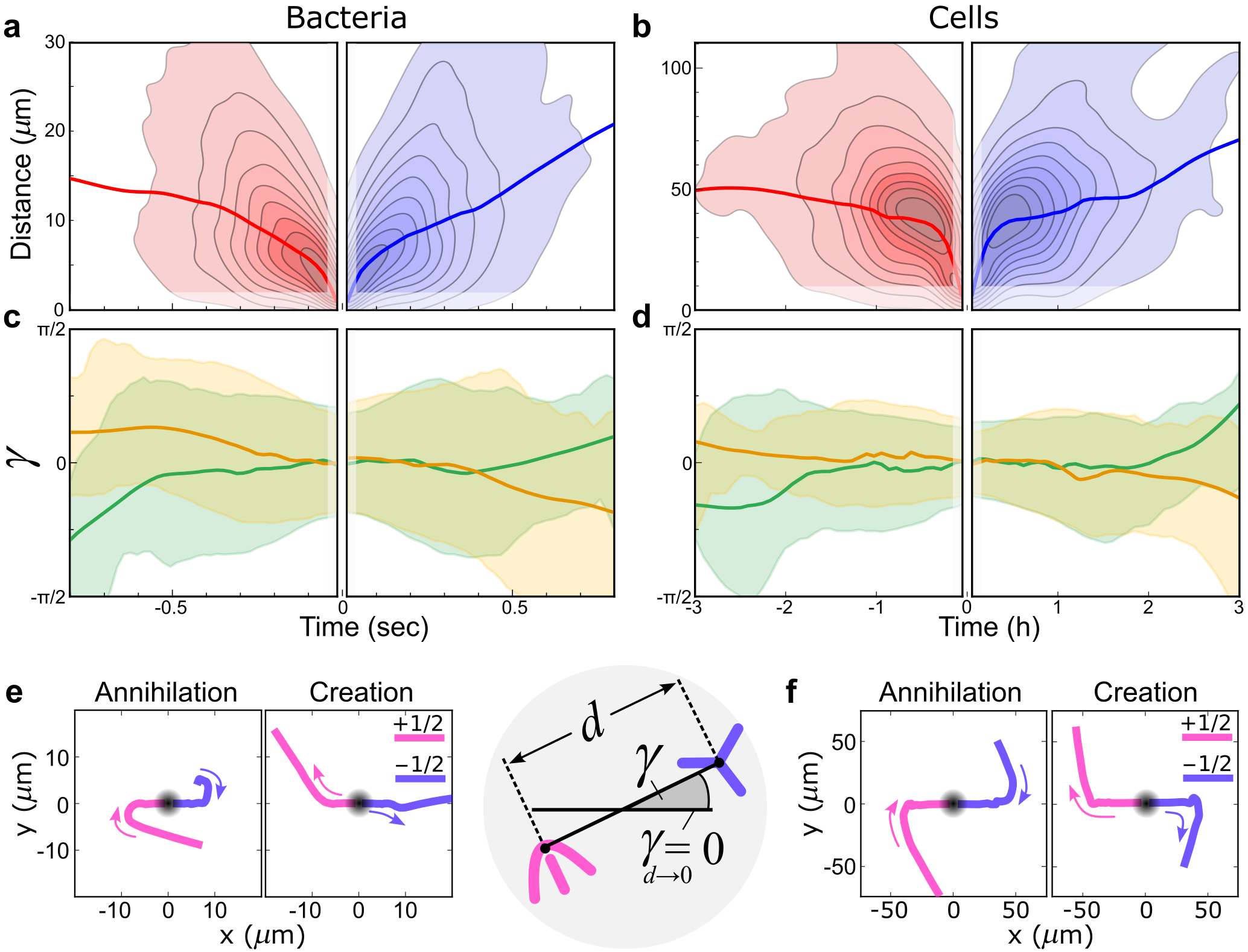} 
        \captionsetup{justification=raggedright} 
	\caption{\textbf{Spiraling trajectories of creation and annihilation.}
    \red{Center sketch: Co-rotation angle $\gamma$—the line connecting between the location of defect centers with respect to the angle close to the creation/annihilation instant ($\gamma_{d \rightarrow 0}$).}
    \textbf{(a, b)} Distance between $\pm^1/_2$ defect-pairs as a function of time after creation (blue) and before annihilation (red), for bacteria and cells. The creation/annihilation instant is indicated by $t({d \rightarrow 0})$, with negative values indicating time to annihilation, and positive ones indicating time from creation. Solid lines represent the average trajectories, while red and blue color intensities indicate probability density. \textbf{(c, d)} Time evolution of $\gamma$ (the angle of the line connecting the defect-pairs, aligned to $\gamma_{d \rightarrow 0}=0$). Solid lines represent averages for ``up'' (green) and ``down'' (yellow) configurations. Shaded areas indicate the standard deviation. \textbf{(e, f)} Average trajectories of $+^1/_2$ (pink) and $-^1/_2$ (violet) defects during creation and annihilation, measured in the lab frame of reference. Trajectories are centered around the creation/annihilation position (black point), rotated to $\gamma_{d \rightarrow 0}=0$ and mirrored to the ``down'' configuration.}
	\label{fig:F3}
\end{figure*}

The defects are created and annihilated in pairs with opposite charges. In both systems, the rates of creation and annihilation events fluctuate over time but stabilizes to a steady state (Figs. \ref{fig:F1}c-d), implying an approximately constant number of defects over time with a total charge that fluctuates around zero. We characterize the motion of defect pairs through these observables, see the sketches in Figs. \ref{fig:F2} and \ref{fig:F3}.
\begin{itemize}
\item $t$: time from creation or annihilation event (negative time indicates time before annihilation).  
\item $d$: the distance between the $+^1/_2$ and $-^1/_2$ defects in the pair.
\item $\gamma$: the argument of the vector from the $+^1/_2$ to the $-^1/_2$ defect core, rotated so that $\gamma(t=0)=0$.
\item $\varphi^+$: the angle between the orientation of the ``tail'' of the comet-shaped $+^1/_2$ defect and the vector connecting the pair. Thus, $\varphi^+ = 0$ indicates that the tail is pointing toward the $-^1/_2$ defect.
\item $\varphi^-$: the angle between the orientation of one of the $-^1/_2$ legs and the vector connecting the pair. We use the smallest angle out of the three legs. For example, $\varphi^- = 0$ implies that one of the legs is pointing away from the $+^1/_2$ defect.
\item $\delta=\varphi^+ -3\varphi^- -\pi$: pair orientation offset. This gives a measure of how compatible the orientations of the defects are with each other. $\delta=0$ suggests the defects are at the orientations that minimize the variation in the nematic field between them~\cite{pearce2021properties,tang2017orientation,vromans2016orientational}.
\end{itemize}

\subsection*{\red{Spontaneous breaking of mirror symmetry during defect creation and annihilation}}

At the point of annihilation or creation, defect pairs are in phase ($\delta = 0$; see Figs.~\ref{fig:F2}a-b), indicating they are perfectly aligned. This is due to elastic torques that diverge when defects are out of phase at small distances~\cite{pearce2021properties}. \red{However, the orientation of the $+^1/_2$ defect relative to the position of the $-^1/_2$ defect (given by $\varphi^+$) is free to vary, provided they remain in phase. In a typical active nematic system, the average orientation of the $+^1/_2$ defect during creation or annihilation is dictated by the active stress, which causes the $+^1/_2$ defect to self-propel; for an extensile active nematic one expects $\varphi^+=0$ during creation and $\varphi^+=\pm\pi$ during annihilation~\cite{giomi2014defect, shankar2018defect}. Conversely, in both our bacteria and cellular experiments, we observe a bimodal distribution of  $\varphi^+$ immediately before annihilation with peaks around $\pm\pi/2$; see Figs.~\ref{fig:F2}c-d.} This indicates that the tail of the positive defect is \red{oriented roughly perpendicular relative} to the direction of annihilation at the end of this process. \red{Since $\delta=0$, $\varphi^-$ varies by about one-third as much.} A similar but less pronounced effect is also visible immediately after creation. This bimodality indicates a spontaneously broken mirror symmetry in the process of creation and annihilation. To quantify this effect, we split all trajectories into two groups based on their time-averaged orientation across the full trajectory, either ``up'' ($\langle\varphi^+\rangle > 0$) or ``down'' ($\langle\varphi^+\rangle < 0$). The orientation $\varphi^-$ follows a similar symmetry breaking, consistent with $\delta=0$ (Figs. \ref{fig:F2}e-g).

\red{
In contrast, active tubulin-kinesin suspensions, which are purely nematic, exhibit no such mirror-symmetry breaking behavior. Analysis of data from Tan et al. \cite{tan2019topological} (Extended Figs. 4c and 5) shows that the distribution of $\varphi^+$ has a single peak around $0$ (creation) or $\pi$ (annihilation), in accordance with the prediction of active nematic theory.
}

\begin{figure}[ht!]
	\centering
	\includegraphics[width=.45\textwidth]{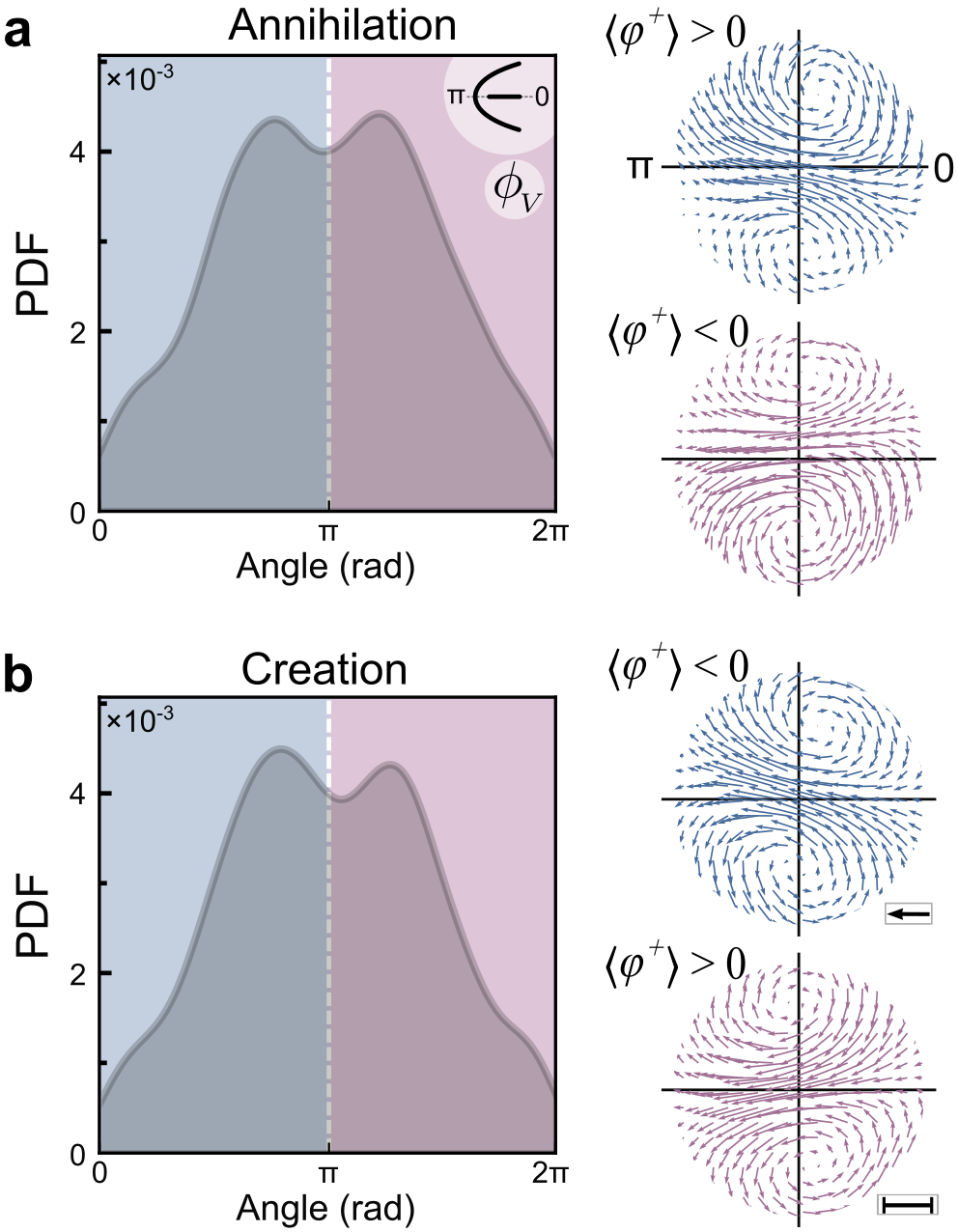} 
        \captionsetup{justification=raggedright} 
	\caption{
    \textbf{Symmetry breaking in the flow around $+^1/_2$.}
    \red{\textbf{(a–b)} The distribution of $\phi_V$ in bacteria, the angle between the $+^1/_2$ velocity and defect orientation (inset in panel a). 
    Average flow fields around $+^1/_2$ defects are obtained by separately averaging trajectories according to their defect-pair configurations: 
    ``up'' ($\langle \varphi^+ \rangle > 0$) or ``down'' ($\langle \varphi^+ \rangle < 0$). Flows are shown for defects with overall displacement to the right (blue--gray) and to the left (pink). 
    The velocity scale indicated by the black arrow is $10~\mu\mathrm{m}/\mathrm{s}$. 
    Scale bar: $10~\mu\mathrm{m}$.
        }
    }
	\label{fig:F4}
\end{figure}

\subsection*{Spiraling trajectories of creation and annihilation}

The separation rate between paired defects (Figs. \ref{fig:F3}a-b) is characterized by increased speed close to the creation or annihilation events, consistent with theoretical predictions for defect annihilation~\cite{giomi2014defect}. However, contrary to typical active nematics, defects do not move in a straight line, but rather spiral toward or away from each other, see Figs. \ref{fig:F3}e-f for average trajectories of defects in the ``down'' configuration. The handedness of the spiral motion can be quantified by plotting $\gamma(t)$; the orientation of the vector connecting the defects. Generally, $\gamma$ varies monotonically with time, indicating that defect pairs tend to co-rotate during the process of annihilation or creation. Furthermore, the direction of the co-rotation is correlated with the ``up'' or ``down'' configuration (Figs.~\ref{fig:F3}c-d). These spiral paths are similar to those predicted for the annihilation of topological defects with an initial phase disparity~\cite{pearce2021properties,vromans2016orientational}, however since we do not observe any major phase disparity (Figs.~\ref{fig:F2}a-b) we predict that the spiraling arises from active forces and self-propulsion of the $+^1/_2$ defect.

\subsection*{\red{Spontaneous breaking of mirror symmetry in the flow fields}}

\red{self-propulsion of $+^1/_2$ defects arises from the characteristic flow fields generated around the defect by the active forces. For a typical active nematic, the flow around a $+^1/_2$ defect features two counter-rotating vortices and a central flux aligned with the defect which causes it to self-propel~\cite{giomi2014defect}.
} 

\red{In order to quantify the direction in which defect move we introduce two additional angles to be measured.

\begin{itemize}
    \item $\phi_V$: For a single $+^1/_2$ defect, $\phi_V$ is the angle between the defect velocity (calculated from the displacement between consecutive frames) and the defect orientation.   
\end{itemize}
In words, $\phi_V$ is the offset between the orientation of a $+^1/_2$ defect and the direction in which it moves. The theory of active nematics predicts that for an isolated defect $\phi_V= \pm \pi$~\cite{giomi2014defect}.
}

\red{Figure \ref{fig:F4} shows the direction of the $+^1/_2$ movement, $\phi_V$, during creation and annihilation events. The distribution of angles is bimodal, suggesting that the spontaneous break in symmetry is also manifested in a slanted trajectory (compared to the defect orientation).
This can be directly observed by averaging the flows around recently created/soon to be annihilated $+^1/_2$'s, partitioned according to their ``up'' or ``down'' configurations (Fig. \ref{fig:F4}).
}

\red{
We find low, yet highly significant correlation between $\phi_V$ close to creation and close to annihilation of the same defect $+^1/_2$ (for details see Extended Data Fig. 6). This indicates that the direction off-axis motion is consistent throughout the lifetime of the defect.}

\red{Overall, the results above suggest that $+^1/_2$ are created with an ``up'' or ``down'' orientation. This orientation persists throughout the life of the defect and finally determines its configuration during annihilation. The configuration of the $-^1/_2$ seems to be constrained or enslaved to the $+^1/_2$, Extended Data Figs. 1 and 2}

\red{Finally, comparing with active nematic tubulin-kinesin suspensions, which do not exhibit mirror-symmetry breaking, the displacement of $+^1/_2$ defects remains aligned with their orientation. This is evidenced by the single peak in the distribution of $\phi_V$ at $\pi$ (Extended Fig. 6). This demonstrates that the mirror-symmetry breaking observed in bacteria and cells does not arise from the presence of counter-rotating vortex structures, which are generic to active nematic turbulence \cite{doostmohammadi2016stabilization}.
}


\begin{figure*}[ht]
	\centering
	\includegraphics[width=.95\textwidth]{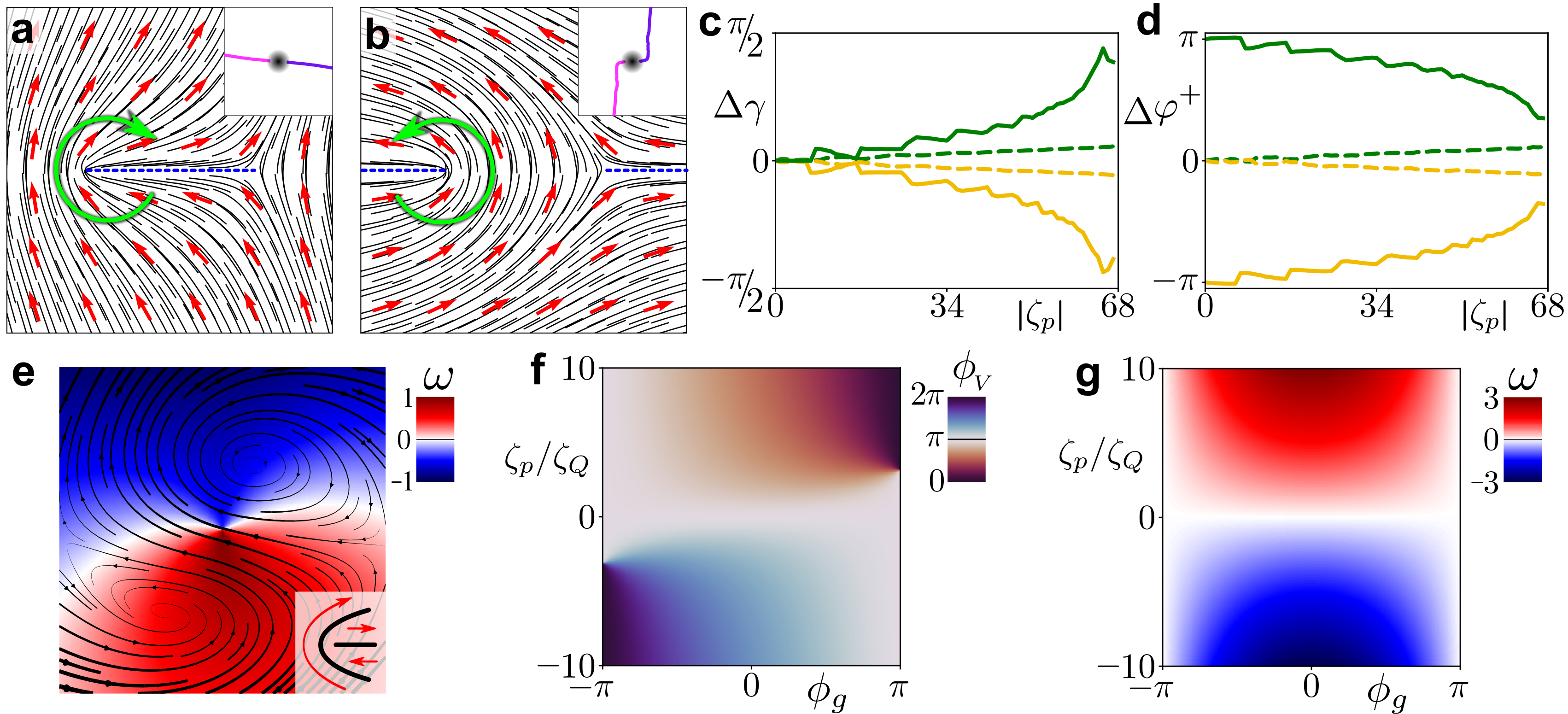} 
        \captionsetup{justification=raggedright} 
	\caption{\textbf{The interaction between nematic and polar symmetries.}
    \red{\textbf{(a)} Defect creation via bend instability with the overlaid nematic (black lines) and polar (red arrows) fields. A grain boundary in the polar field is introduced (blue dashed line). The polar field leads to a torque on the defects (green arrow), which will reorient the $+^1/_2$ defect, causing it to self-propel and spiral around the $-^1/_2$ defect (inset). Shown here is the ``down'' configuration. $\zeta_p\neq0$. \textbf{(b)} Nematic and polar fields around a pair of defects before annihilation. Here, the grain boundaries do not directly connect. A similar torque, acting on the $+^1/_2$ defect, leads to a spiraling trajectory (inset). \textbf{(c)} Total co-rotation ($\Delta\gamma$) and \textbf{(d)} final positive defect orientation ($\Delta\varphi^+$) of the defect pair as a function of the applied polar body force ($\zeta_p$). Dashed lines indicate creation and solid lines for annihilation, green ``up'' and yellow ``down'' configuration. \textbf{(e)} Predicted Stokes flow around a $+^1/_2$ defect in the annihilation ``up'' (or creation ``down'') configuration. Shown with $|\zeta_p/\zeta_Q| = 0.5$ and $\phi_g=0$. \textbf{(f)} Angular offset of defect self-propulsion ($\phi_V$) and \textbf{(g)} net vorticity around a $+^1/_2$ defect as a function of the relative polar body force ($\zeta_p/\zeta_Q$) and position of the grain boundary ($\phi_g$). $\phi_g=0$ implies that the grain boundary alignment with the tail of the $+^1/_2$ defect is as in \textbf{(a)} and \textbf{(b)}.}}
	\label{fig:F5}
\end{figure*}

\subsection*{The source of mirror symmetry breaking}
\red{In this section we provide a possible physical explanation to the symmetry breaking observed above. Cellular systems such as those studied here are often analyzed using the framework of active nematics with good agreement. In particular, the constant turnover of half integer defects, in which $+^1/_2$ defects are created through bend instability and self-propulsion, indicates that the system behaves as an extensile active nematic. However, in addition to nematic order, the bacteria and cells studied here also exhibit polar symmetry through their directed self-propulsion mechanism \cite{zhou2014living,be2019statistical,vaidvziulyte2022persistent,lo2024spontaneous,lacroix2024emergence}. Combining nematic interactions with active stresses of various symmetries can significantly alter active nematic behavior \cite{patelli2019understanding,ardavseva2025beyond, grossmann2020particle,de2025self}. In particular, breaking the polar symmetry of the nematic field at the microscopic level can lead to a break in chiral symmetry at the macroscopic level.}

\red{To describe the orientation of the cells we utilize a nematic tensor field, $\underline{\underline{Q}}$, which governs the steric interactions between the cells. In addition, we describe the head-tail symmetry of the cells by a polar vector field, $\underline{p}$, which governs the self-propulsion of the cells against the substrate. This can be visualized as two superimposed parallel fields: one with lines describing the long axes of the cells, and another with arrows for the head direction (see Fig.~\ref{fig:F5}a). Both $\underline{p}$ and $-\underline{p}$ are compatible with the same nematic field.}

\red{We model the system using an active nemato-polar framework in which we simultaneously simulate $\underline{\underline{Q}}$ and $\underline{p}$. The dynamics of $\underline{p}$ are constrained to follow the dynamics of $\underline{\underline{Q}}$ keeping the two fields parallel. To generate the motion of the cells we include two active terms: a standard extensile active stress, $\sigma^a = \zeta_Q\underline{\underline{Q}}$, and a body force describing cellular self-propulsion against the substrate,  $\underline{f} = \zeta_p\underline{p}$.
See SI for details.}

\red{In a typical extensile active nematic, defects are created when a bend distortion in the director field becomes unstable. The bend distortions are characterized by a non-zero curl, which is unsigned for the nematic field. However, in the polar field the bend distortion is signed and therefore breaks mirror symmetry. The resulting half integer defects are incompatible with the polar field. This leads to a grain boundary that connects the two defects, across which the polar field changes sign, see blue line on Fig.~\ref{fig:F5}a. Notably, the mirror image of this process produces defects with opposite chirality and is equally likely, indicating that defect chirality selection is driven by spontaneous symmetry breaking during creation. In this context, distinct ``up'' and ``down'' defect configurations can emerge during creation.}

\red{We simulate the defect creation process by initializing the fields with a single prominent bend distortion at the center of the simulation domain. Upon simulation, the bend distortion nucleates a pair of defects that separate and follow curved trajectories, see Fig.~\ref{fig:F5}a (inset), recreating those observed in experiments, see Fig.~\ref{fig:F3}e-f. We simulate the annihilation process by initializing the simulation with a pair of defects such that $\varphi^+ = \pm \pi$, (see Fig.~\ref{fig:F5}b), as expected in active nematic.  
Simulated annihilations also show curved trajectories (Fig.~\ref{fig:F5}b inset), resembling the experimental observations (Fig.~\ref{fig:F3}e-f). We quantify the co-rotation of the defect pair by plotting the total change in $\gamma$ throughout the creation or annihilation process, see Fig.~\ref{fig:F5}c. We see that the degree of co-rotation increases with the strength of the polar forcing, $\zeta_p$, and is larger during annihilation. We also observe rotation of the $+^1/_2$ defect, quantified by $\Delta\varphi^+$ at the end of the simulation, shown in Fig.~\ref{fig:F5}d. The rotation of the defect increases with the magnitude of the polar force. Consistent with our experimental observations (Fig. \ref{fig:F3}e–f), the rotation is more pronounced during annihilation than during creation. Importantly, the chirality of the simulations can be reversed either by mirroring the fields or, equivalently, by changing $\zeta_{p}\rightarrow-\zeta_{p}$. This procedure allows to simulate the system in either the ``up'' or ``down'' configuration, shown as green or yellow in Figs.~\ref{fig:F5}e-f. If we remove the active body force, $\zeta_p = 0$, we recover the straight trajectories expected for typical active nematics.}

\red{In our experiments, the flow profile around a $+^1/_2$ defect is offset relative to the symmetry axis of the defect, see Fig.~\ref{fig:F4}. We can analytically calculate how the introduction of the polar body force influences the flow around a $+^1/_2$ defect and observe a similar offset, see Fig.~\ref{fig:F5}e and SI for details. The polar body force breaks the mirror symmetry of the defect, resulting in a central flux at an angle $\phi_V$ relative to the tail of the $+^1/_2$ defect. The magnitude of this angular offset depends on the angular position of the grain boundary ($\phi_g$) and the scale of the polar active force, see Fig.~\ref{fig:F5}f. The distribution of $\phi_V$ measured from our experiments has a peak at $\approx\pm\pi/4$ for bacteria and $\approx\pm\pi/3$ for cells, see Fig. \ref{fig:F4} and Extended Fig. 6a-b.
Such an offset would indicate that the polar forcing accounts for at least $25\%$ of the flow around the defect. The polar body force also introduces a net vorticity around the core of the defect, see Fig.~\ref{fig:F5}g. While the net flux at the core of the defect causes it to self-propel, the net vorticity induces a torque that causes defects to rotate. Similar to the velocity offset, the magnitude of the rotation is controlled by the magnitude of the polar active force, but the direction of rotation depends only on the sign of $\zeta_{p}$, hence the chirality of the defect.}

\red{Taken together, these observations explain the spontaneous symmetry breaking into ``up'' and ``down'' configurations observed in experiments. Defects are generated through a bend instability due to the extensile active stress, hence with $\varphi^+\approx0$. This process breaks the mirror symmetry of the polar field and introduces a grain boundary across which $\underline{p}$ changes sign (blue dashed line in Fig.~\ref{fig:F5}a). The grain boundary connects the defects and results in a net curl around the $+^1/_2$ defect which translates to a net torque (green arrow in Fig.~\ref{fig:F5}a). This torque causes the $+^1/_2$ defect to rotate as it self-propels, thus $\dot{\varphi}^+\neq0$. Combined with the defect self-propulsion, this rotation results in the defect taking a curved trajectory, spiraling away from its $-^1/_2$ equivalent (Fig.~\ref{fig:F5}a). The sign of the torque depends on the sign of the curl of the initial bend distortion. Thus, it arises from a spontaneous break in symmetry. A counter clockwise net torque on the defect leads to $\dot{\varphi}^+>0$ and we observe the ``up'' configuration, as shown in Fig.~\ref{fig:F5}a.}

\red{The polar grain boundary connected to the defect core is topologically constrained and persists throughout the defect's lifetime. Thus, a $+^1/_2$ defect has an offset flow profile throughout with a chirality that is determined at creation (Fig.~\ref{fig:F5}e). The latter drives the bimodal distribution of $+^1/_2$ defect velocities as ``up'' and ``down'' defects self-propel with an angular offset (Figs.~\ref{fig:F4},~\ref{fig:F5}f, and Extended Fig. 6. This indicates that the chiral symmetry of a $+^1/_2$ is broken at creation and persists until annihilation, where it manifests again in that process.}

\red{During annihilation, the $+^1/_2$ defect tail initially points away from the $-^1/_2$ defect, hence $\varphi^+=\pm\pi$. In this case a clockwise net torque around the $+^1/_2$ defect results in $\dot{\varphi}^+<0$ which causes the defect to rotate into the ``up'' annihilation configuration $\varphi^+>0$ (Fig.~\ref{fig:F5}b). The latter is the chirality associated with defects that originated in the ``down'' configuration at creation. Hence, we expect the chirality of $+^1/_2$ defects created in the ``up'' configuration to be identical to those that annihilate in the ``down'' configuration, as indeed observed in experiments (Fig.~\ref{fig:F2}). As in creation, this rotation, combines with defect self-propulsion, results in a curved trajectory (Fig.~\ref{fig:F5}d). In contrast to the creation process, annihilation is independent of $\varphi^+$ (assuming $\delta = 0$). Thus, defects can annihilate even when $\varphi^+\neq 0$, as shown in Figs.~\ref{fig:F2}c-d. }

\subsection*{Time-reversal symmetry breaking}
Active systems are typically irreversible, which is manifested in strictly positive entropy production (EP). Loosly speaking, EP is defined as the Kullback-Leibler divergence (KLD) from the distribution of trajectories compared to the distribution of time-reversed trajectories~\cite{seifert2012stochastic,fodor2022irreversibility}. In some systems, such as Langevin particles satisfying the Einstein relation, EP equals (in units of temperature) to the dissipated heat~\cite{fodor2022irreversibility,ro2022model}. It has been suggested that EP also plays a role in active systems, and is indicative of the amount of extractable work~\cite{fodor2022irreversibility,ro2022model}. Previous attempts to quantify EP in bacterial and cellular dynamics proved challenging. In~\cite{ro2022model}, local EP in swimming bacteria was estimated using a compression-based methods, and found positive EP mainly close to walls or barriers. However, negligible or no EP was observed in the bulk, despite the fact that the system is highly active and out of equilibrium.

Obtaining accurate statistics of the joint distribution of cellular or bacterial trajectories is challenging because of the high dimensionality of such a dataset as estimating KLD suffers from severe undersampling of the phase space~\cite{ro2022model}. 
Here, defects provide physically motivated coarse-graining of the micro-scale  motion into a meso-scopically observable feature. As a consequence, all the estimates reported here are lower bounds to the actual EP. 

In extensile-active nematics, $+^1/_2$ defects move in a direction that is opposite to the ``comet-tail''~\cite{giomi2015geometry}. To quantify this effect, Table \ref{tab:table1} shows the average projection of the defect direction onto the velocity, which is indeed negative. We can calculate a similar measure by contracting the velocity of a $-^1/_2$ defect with the third rank tensor that describes its orientation (see Methods). Interestingly, we find a non-zero average value (Table \ref{tab:table1}) that is not expected in a typical active nematic. This is consistent with the net force we calculate on the core of a $-^1/_2$ defect resulting from the polar active forcing (see SI). 

Table \red{\ref{tab:table1}} also shows results for EP estimates. For bacteria, we find a high value of EP in $+^1/_2$ defects, and lower (but statistically significant) for $-^1/_2$ ones. For cells, the EP in $+^1/_2$ cells is an order of magnitude smaller than for bacteria. For $-^1/_2$ it is below the error range. 

Next, we turn to analyze creation and annihilation events. Figures \ref{fig:F2}c-d show that the distribution of $\varphi^+$ is different close to creation compared to annihilation. Similar broken symmetries can also be observed in the average trajectories (Figs. \ref{fig:F3}e-f) and the flow fields (Extended Data Figs. 1 and 2).

Focusing on time-reversal, running the experimental movies backward, creation will appear as annihilation and vice versa. To this end, Table \ref{tab:table2}, column labeled cross, shows the symmetric KL divergence (Jeffreys divergence) between forward-creation/backward-annihilation and forward-annihilation/backward-creation,
\begin{equation}
\begin{aligned}
   \red{\frac12 D_{\rm KL} ( \overrightarrow{\rm creation} || } & \red{\overleftarrow{\rm annihilation} ) +} \\
   & \red{\frac12 D_{\rm KL} ( \overrightarrow{\rm annihilation} || \overleftarrow{\rm creation} ) } .
\end{aligned}
\end{equation}
The positive symmetric divergence indicated that creation and annihilation are not time-symmetric, both for bacteria and cells.

Comparing the total EP rates, we need to consider the rate of creation and annihilation events. For bacteria, the time between events is about 1 sec (50 frames), hence the total EP due to movement is about 50 times larger than due creation-annihilation. For cells, the time between creation and annihilation is about 2 hrs (24 frames). Hence, EP due to movement is comparable to EP due to creation-annihilation. These differences are consistent with the observation that in bacteria, $+^1/_2$ defects appear self-propelled in the sense that their trajectories are ballistic over short time periods \cite{yashunsky2024topological}, implying a relatively large EP. With cells, both $\pm^1/_2$ defects are diffusive (or the timescale of ballistic motion is relatively much shorter) \cite{blanch2018turbulent}, their movement in the bulk is close to reversible, and EP is most apparent in creation and annihilation.

\begin{table*}
    \centering
    \caption{\textbf{Entropy production (EP) estimation for isolated defects.} The row noted projection shows the projection of the defect orientation on the velocity. The row noted EP shows a statistical estimate for a lower bound for the entropy production rate obtained for defect trajectories. Values show the estimate $\pm$ the standard error. Since this is a lower bound, a value that is statistically larger than zero implies non reversible dynamics. See the methods section for details.}
    \label{tab:table1} 
    \begin{tabular}{c|c|c|c|c}
        \hline
        \multicolumn{1}{c|} { }& \multicolumn{2}{c|}{\textbf{Bacteria}} & \multicolumn{2}{c}{\textbf{Cells}} \\
        \multicolumn{1}{c|}{Defect charge} & $+^1/_2$ & $-^1/_2$ & $+^1/_2$ & $-^1/_2$ \\
        \hline
        projection   & $-0.2\pm 0.003$ & $0.005 \pm 0.003$ & $-0.08\pm 0.003$ & $0.01\pm 0.003$ \\
        EP    & \red{$0.94\pm 0.03$} & \red{$0.005\pm 0.003$} & \red{$0.18\pm 0.01$} & \red{$0.0006\pm0.0007$} \\
        \hline
    \end{tabular}
\end{table*}

\begin{table*}
    \centering
    \caption{\textbf{Entropy production (EP) estimation for defect pairs.}
    The table shows different statistical estimates for a lower bound for the entropy production rate obtained for defect trajectories. The dimension $D$ indicated the length of the trajectories used (starting at the event). Values show the estimate $\pm$ the standard error. Since this is a lower bound, a value that is statistically larger than zero implies non reversible dynamics. See the methods section for details.
    }
    \label{tab:table2} 
    \begin{tabular}{c|c}
        \hline
        \textbf{Bacteria} & \textbf{Cells} \\
        \hline
        \red{$0.05\pm0.01$} & \red{$0.5\pm0.2$} \\
        \hline        
    \end{tabular}
\end{table*}

\subsection*{Conclusions}
\red{By studying defect dynamics in two distinct active living systems—swarming \textit{Bacillus subtilis} bacteria and human bronchial epithelial cell (HBEC) monolayers—we demonstrate that topological defect creation and annihilation are fundamental, symmetry-breaking processes that drive these systems far from equilibrium. Despite orders-of-magnitude differences in timescales and propulsion mechanisms, both systems exhibit nematic half-integer defects whose pair dynamics break both mirror symmetry and time-reversal symmetry. Moreover, $+^1/_2$ defects spontaneously adopt persistent “up” or “down” chiral configurations and follow spiral trajectories.
}

\red{\subsubsection*{Irreversibility and Entropy Production}} \red{The non-equilibrium and irreversible nature of active living systems is unequivocal.
Because bacteria and cells are self-propelled, the dynamics is out of equilibrium even at the level of the individual. Therefore, the theoretical prediction is that irreversibility should be observed at the single-cell level.  
However, the difficulty in capturing irreversibility from individual trajectories is intuitively clear. For any observed single-cell trajectory, one can imagine a cell tracing the reversed path. Thus, EP from single cells is only expected to be captured by adding intra-cellular degrees of freedom corresponding to chemical and biological function or the underlying self-propulsion mechanism.
From this respect, defects are physically intuitive, virtual meso-scopic degrees of freedom. Virtual in the sense that they are not the particles of which the system is made of. Meso-scopic since each one is defined by a small local collective of several close cells.

Our findings thus show that irreversibility is statistically evident locally at the level of the collective; see also \cite{tan2019topological} for measurements of topological entropy and \cite{radhakrishnan2025irreversibility} for a theoretical study. 
Moreover, we find that defect creation and annihilation are not reciprocal processes and that they generate entropy. This is also physically intuitive as the continuous creation of defects in nematic systems is a hallmark of activity.
In both experimental systems, entropy production during defect creation and annihilation accounts for a disproportionately large fraction of the total entropy production. In cells, it is an order of magnitude higher, suggesting that these events are key drivers of non-equilibrium dynamics rather than incidental fluctuations. Still, one should remember that our estimates are merely lower bounds for the actual EP, which means that quantitative comparison is still lacking.

Overall, entropy production estimates bring forth some fundamental differences between swarming bacteria and crawling cells regarding the main sources of entropy production (defect motion in bacteria and creation-annihilation in cells). This points to the physical mechanisms underlying departure from equilibrium.
}

\red{\subsubsection*{Spontaneous Symmetry Breaking and the Polar-Nematic Model}

Our experiments reveal that topological defect creation and annihilation in living active nematics spontaneously break mirror symmetry—a phenomenon absent in standard active nematic theory. Specifically, $+^{1}/_{2}$ defects near a $-^{1}/_{2}$ defect adopt one of two distinct chiral configurations, exhibiting bimodal orientation distributions and persistent spiral trajectories. This chiral symmetry breaking is established during defect creation and persists until annihilation.

To explain these observations, we introduce an active nemato-polar model that couples a polar vector field $\bm{p}$—representing head–tail asymmetric self-propulsion—to the nematic tensor order parameter $\bm{Q}$. A half-integer topological defect in the nematic field imposes a $\pi$-discontinuity (grain boundary) in the polar field, breaking the mirror symmetry of the $+^{1}/_{2}$ defect core. This generates a net active torque that rotates the defect as it self-propels. The torque sign, selected spontaneously during nucleation, determines the chirality and drives the spiral motion of defect pairs.

The model reproduces key experimental signatures: the bimodal distribution of orientation angle $\varphi^{+}$, the chiral rotation quantified by $\Delta\gamma$, and the flow field offset around $+^{1}/_{2}$ defects. It also explains the stark contrast with achiral active nematics (e.g., microtubule–kinesin suspensions), where the absence of a polar field yields dynamics consistent with standard nematic theory.

We suggest that spontaneous symmetry breaking in living active matter stems directly from the intrinsic polar activity of its constituents. A complete description therefore requires accounting for the interplay between nematic alignment and polar self-propulsion, revealing how microscopic asymmetry is imprinted onto meso-scopic topological degrees of freedom.
} 

Overall, our results challenge the standard active nematic description often applied to living active matter. By uncovering these universal physical principles, we advance understanding of biological self-organization, revealing the fundamental role of symmetry in shaping life at multiple scales.



\section{Methods}
\subsubsection*{Swarming bacteria}
Experiments were conducted on \textit{Bacillus subtilis} 3610, the ``wild-type'' (WT) strain, which is a rod-shaped flagellated species with dimensions $1 ~\mu m \times 7 ~\mu m$. Cells are stored at $-80~^{\circ}C$ in frozen stocks. Fluorescently labeled variants, green and red, are used 
(\textit{amyE::PvegR0\_sfGFP;specR} and \textit{amyE::Pveg\_R0\_mKate;specR}); 
labeling does not affect any known measured quantity. In order to resolve the cells in the crowded colony, we mix the two labeled strains at ratio $1:1$ (after separate overnight growth), which yields much sparser fields of view (see camera settings next). Swarm experiments are typically done in a standard Petri-dish ($8.8 ~cm$ in diameter), where the colonies are grown on soft agar ($0.5 ~\%$ and $25 ~g/L$ LB) and aged in ambient lab conditions ($24~^{\circ}C$ and $35~\%$ RH) for 4 days. The latter forms the thicker colony structure compared to results obtained in other past studies. Overnight cultures are grown from isolated colonies (which are grown from frozen stocks on hard agar plates ($2~\%$ and LB)) at shaking ($200$ rpm and $30~^{\circ}C$) for $18 ~h$ in LB liquid medium ($2 ~mL$ in a $15 ~mL$ tube). A small drop of overnight culture ($4 ~\mu L$), either from a single strain, or a $1:1$ mix of the two labeled variants, is deposited on the agar in the middle of the plate. The colonies grow for a few hours in a $95~\%$ RH incubator at $30~^{\circ}C$. The swarm colonies form a quasi-2D structure with a thickness of $\sim 7 ~\mu m$, which is uniform along centimeter-wide distances. Observations are done using an optical microscope (Axio Imager Z2 Zeiss; 40×) operated at fluorescence mode; this magnification yields a $150 ~\mu m \times 150 ~\mu m$ observation frame. The system is equipped with a splitting device (Optosplit II) that enables dual excitation and acquisition. The two fluorescence fields are obtained for the same spatial field and time. The image is projected on a NEO camera with $1800 \times 900$ pixels and at 50 frames per second so that the green cells are seen on the left panel while the red ones are seen on the right panel. In the current study we focus on the upper layer of the colony. The system operates with the dual fluorescence set Ex 59026x, beam splitter 69008bs, and Em 535/30; 632/60. A compensation lens (an integral part of the Optosplit II device) is used to adjust the focus in each of the panels so that both the green panel and the red one are in focus at the upper layer. 

\subsubsection*{Human bronchial epithelial cells (HBECs)}
Human bronchial epithelial cells (HBECs) cultured as previously reported. HBEC culture was executed in a supplemented keratinocyte serum-free medium with l-glutamine (Keratinocyte-SFM with l-glutamine; Gibco). Cells were maintained at $37~^{\circ}C$ under $5~\%$ $CO_2$ partial pressure and $95~\%$ relative humidity. For time-lapse experiments, cells were seeded at a density of approximately 640,000 cells per well in polystyrene (TPP) 12-well tissue culture plates.
Time-lapse microscopy experiments were performed in phase contrast on an automated IX71 inverted microscope (Olympus) equipped with the same temperature, humidity, and $CO_2$ regulations as in culture (Life Imaging Services). Images were acquired using a $10\times$ objective, providing a $1.5 ~mm \times 1.5 ~mm$ observation frame. The intervals between images were set to 5 minutes.

\subsubsection*{Measurement of Velocity and Director Fields}
Image analysis was conducted using custom Python code. To obtain the velocity field from microscopic images, we employed the Gunnar Farneback method available from OpenCV library producing displacement vectors for every pixel rather than a sparse set of points \cite{farneback2003two}. The director field was computed using the second-moment matrix, implemented with the structure tensor function from the skimage Python library. Both the flow and director fields were computed with an averaging window size of $2.5 ~\mu m \times 2.5 ~\mu m$ for bacteria and $11.1 ~\mu m \times 11.1 ~\mu m$ for cells.

\subsubsection*{Defect Detection and Tracking}
\red{Topological defects were identified by evaluating the winding of the director field. We computed the topological charge $k$ at each grid point by summing the angular differences of the director along a closed loop around a $2 \times 2$ unit cell. Because the nematic phase is apolar, all angular differences were wrapped to the interval $(-\pi/2, \pi/2]$. Grid points with a net winding of $\pm \pi$ were labeled as defects with strength $k=\pm ^1/_2$.

Defect orientations were determined using a direct interpolation method. The local director field was interpolated onto a circular path centered on each defect. We then extracted the orientation angle by maximizing the agreement between this interpolated profile and the analytical director field for a defect of strength $k$. For $+^1/_2$ defects, this yielded a single orientation axis that points toward the comet head. For $-^1/_2$ defects, the method resolved the three-fold symmetry of the core and returned the three equivalent orientation angles separated by $120^\circ$.

Defect trajectories were tracked using the trackpy Python library \cite{crocker1996methods}. The tracking results were merged with the orientation data to produce a list of defects with their charges, orientations, and positions.
}
Overall, the data on a single $+^1/_2$ defect trajectory is a sequence of vectors $\textbf{p}_0,...,\textbf{p}_K$, where each vector pi is in $R^3$, i.e., each frame consists of 3 scalar values, corresponding to the $x-y$ coordinates and an angle, corresponding to the orientation of the $+^1/_2$ defect. The data on a single $-^1/_2$ defect trajectory is a sequence of vectors $\textbf{n}_0,...,\textbf{n}_K$, where each vector $\textbf{n}_i$ is in $R^5$, i.e., each frame consists of 5 scalar values, corresponding to the x-y coordinates and the three angles, corresponding to the three directions of the $-^1/_2$ defect. They may not be exactly $2\pi/3$ apart. Velocities were taken as the displacements between consecutive frames with no further smoothing.

\subsubsection*{Defect-Pair Detection}
To identify defect-pairs, thresholds for distance ($\Delta d$) and time ($\Delta t$) were applied. The thresholds were set at $\Delta d = 2 ~\mu m$ for bacteria and $\Delta d = 11.1 ~\mu m$ for cells, with \red{$\Delta t = 0.04$ seconds for bacteria and 10 minutes for cells}. Two oppositely charged defects, $a(t_a,~x_a,~y_a)$  and $b(t_b,~x_b,~y_b)$ were considered a created pair if their initial detection times, $t_a$ and $t_b$, satisfied $|t_a - t_b| < \Delta t$, and their initial locations, $x_a,~ y_a$ and $x_b,~y_b$, satisfied $((x_a-x_b)^2 + (y_a-y_b)^2)^{1/2} < \Delta d$. Similarly, oppositely charged defects were identified as annihilation pairs if their final detection times and locations met the same criteria. \red{The pair detection algorithm is designed to prevent more than one annihilation or creation event for any defect. Each defect's trajectory is assigned a unique identifier, ensuring that the same identifier cannot be used in more than one pair. Trajectories that span fewer than five consecutive frames were excluded from the analysis.} Events near the edge of the frame were excluded to prevent interference from off-frame defects. Additionally, defects appearing at the start or disappearing at the end of the time sequence were not considered created or annihilated to ensure accuracy. The detection algorithm identified $1679$ creation and $1676$ annihilation events in bacteria \red{from three experimental replicas}, and $1201$ creation and $1187$ annihilation events in HBECs \red{from six experimental replicas}.

Overall, the data on a single defect-pair is a sequence of vectors $\textbf{x}_0,...,\textbf{x}_K$, where each vector $\textbf{x}_i$ is in $R^8$ ($x,y$ coordinates of each defect, one angle for $+^1/_2$ and three angles for $-^1/_2$). For a creation event, $\textbf{x}_0$ is the measurement closest to creation. For an annihilation event, $\textbf{x}_K$ is the measurement closest to annihilation. Furthermore, each pair-trajectory is \red{classified } as either “up” or “down”, based on the average projection of the $+^1/_2$ direction on the direction connecting the pair \red{(see sketch, Fig. \ref{fig:F2})}. \red{Specifically $\langle \varphi^+ \rangle >0$ was defined as ``up'' and $\langle \varphi^+ \rangle <0$ as ``down'', where $\langle \cdot \rangle$ denotes the circular mean over the trajectory of the paired $+^1/_2$ defect.} 

\subsubsection*{Trajectories of Creation and Annihilation}
To describe the average trajectory of creation and annihilation events, consider the $\textbf{k}$’th trajectory with defined positions. The distance between defect-pairs is calculated as the Euclidean distance between the defect centers, $d=((x_a-x_b)^2 + (y_a-y_b)^2)^{1/2}$, where \textit{a} and \textit{b} refer to the to $+^1/_2$ and $-^1/_2$ defects, respectively. The angle of each defect, $\theta_{a/b}(t)$, was measured at various points along its trajectory relative to the position of the creation/annihilation event ($x_0,y_0$). Since ``up'' and ``down'' defects rotate in opposite directions, we analyzed and averaged the trajectories of the ``up'' and ``down'' configurations separately. Finally, we combined the ``up'' and ``down'' trajectories by transforming the averaged ``down'' trajectory $\langle \theta_{a/b}(t) \rangle$ into $\langle -\theta_{a/b}(t) \rangle$.

\subsubsection*{Average Director and Flow Fields Around Defects}
To measure the average fields around $\pm^1/_2$ defects, we analyzed all defect-pairs based on their configuration (``up'' or ``down'') and trajectory type (``creation'' or ``annihilation''), resulting in four distinct groups. Defects were analyzed alongside their corresponding flow fields within a window of $35 ~\mu m \times 35 ~\mu m$ for bacteria and $300 ~\mu m \times 300 ~\mu m$ for cells, centered at the defect core. The local director and velocity fields were then rotated to align all defects in the same orientation. Finally, the averaged, aligned fields were calculated for each defect type and group.

\subsubsection*{Quantifying irreversibility in single-defect trajectories}

Here, we provide details for the statistical estimates presented in table~\red{\ref{tab:table1}}.

For each $+^1/_2$ defect, the detection algorithm described above provides a sequences of positions in 2D and orientation angle, $\theta \in [0,2\pi)$. The defect velocity $v$ is obtained from the displacements between consecutive frames (next frame - current frame). Denoting the velocity direction $\hat{V} = v /|v|$ and vectorial orientation $\hat{n} = (\cos \theta,\sin \theta)$, the projection of the defect orientation on the velocity direction (row 1 in  table~\ref{tab:table1}) is $\hat{V} \cdot \hat{n}$.  
The table shows the mean over all defects $\pm$ the standard error.

For each $-^1/_2$ defect, the detection algorithm described above provides a sequences of positions in 2D and three orientation angles, corresponding to the directions of the three legs $\theta_1,\theta_2,\theta_3 \in [0,2\pi)$.
We define the average defect direction (taking into account the rank three tensor symmetry) as $p=\sum_i (cos 3 \theta_i, \sin 3 \theta_i)$. The angle corresponding to the defect orientation is then $\varphi = \tan^{-1} (p_y/p_x)/3$. Note that a function such as atan2 should be used to obtain an angle in $[0,2\pi)$. 
Finally, the projection of defect orientation in the velocity direction (angle $\psi$) is given by $\cos 3 (\varphi-\psi)$.
The table shows the mean over all defects $\pm$ the standard error.

The informatic entropy production is the Kullback-Leibler divergence between the distribution of forward observations $\textbf{f}_i$ and backward ones $\textbf{b}_i$. Hence, our experimental data provides us with samples from both spaces. In backward trajectories, the sequence of positions is inverted, and the velocity changes sign (because velocities are a forward difference rule). However, defect angles do not change; hence the projection also changes sign, hence the distribution of forward projections and backward ones differ. Cutting trajectories into 2-frame snippets, we obtain a coarse-grained picture, which is a lower bound of the actual EP.

To estimate EP, forward trajectories use the same single-frame projections described above. For backward trajectories, each trajectory is reversed and analyzed similarly. 
\red{For example, to obtain $+1/2$ trajectories of length $d$, we look at frame segments of length $d+1$. The data for such a segment consists of $(x_1,y_1,
\theta_1),\dots,(x_{d+1},y_{d+1},
\theta_{d+1})$, where $(x,y)$ are the center of mass positions and $\theta$ the defect orientation.
The forward velocities are $v_i^{\rm F}=(x_{i+1}-x_i,y_{i+1}-y_i)$, and the forward orientations, $\hat{p}_i^{\rm F} = (\cos \theta_i,\sin \theta_i)$, $i=1\dots d$. Then, the forward projections trajectory is $\vec{F}=(\hat{v}^{\rm F}_1\cdot\hat{p}^{\rm F}_1,\dots,\hat{v}^{\rm F}_d\cdot\hat{p}^{\rm F}_d)$, with $\hat{v}^{\rm F}_i=v^{\rm F}_i/||v^{\rm F}_i||$.
Similarly, the backward trajectory is $\vec{B}=(\hat{v}^{\rm B}_1\cdot\hat{p}^{\rm B}_1,\dots,\hat{v}^{\rm B}_d\cdot\hat{p}^{\rm B}_d)$, with
$v_i^{\rm B}=(x_{i}-x_{i+1},y_i-y_{i+1})=-v_i^{\rm F}$, and the  orientations, $\hat{p}_i^{\rm B} = (\cos \theta_{i+1},\sin \theta_{i+1})$, $i=1\dots d$. Note that backward orientations are shifted by one frames, compared to the forward one. Trajectories for  $-^1/_2$ defects are computed similarly, taking into account the three-fold symmetry as described above.
}

See below for the estimation method for Kullaback-Leibler divergence.

\subsubsection*{Quantifying irreversibility in defect pairs}
\red{There are 4 angles associated with the creation/annihilation processes: there related to the defect orientation -  $\varphi^+$, $\varphi^-$ and $\delta$ - and one related to the line connecting the defect centers, $\gamma$.
We have defined $\gamma$ relative to the angle between the $\pm^1/_2$ defects at the time of the event. Hence, it is not a state function (it cannot be obtained from a single frame). Hence, using it may introduce irreversibility, because its measurement is with respect to a fixed event which is either set in the past or future. Similarly, $\varphi^+$ and $\varphi^-$ are relative to $\gamma$.
However, both $\delta$ and $\Delta \varphi = \varphi^+ - \varphi^-$ are not (here, $\varphi^-$ is the angle of the leg which is closes to the comet-tail of the +). Therefore, one should use these angles when looking for irreversibility.}

\red{At each event, creation or annihilation, $\delta$ approaches zero. Hence, this sense, creation and annihilation are time-symmetric. As we demonstrate below, the statistics of how $\delta \to 0$ may be more subtle. However, substituting $\delta=0$, $\Delta \varphi = 2 \varphi^+ / 3 + \pi/3$, which is not expected to be reversible.}

\red{
More specifically, for each event we look at a short sequence of $D$ snapshots (before annihilation or after creation). Looking at $\delta$ and $\Delta \varphi$, a forward trajectory yield a $2D$ dimensional vector  $\textbf{f}_i=(\delta_1,\Delta \varphi_1, \dots , \delta_D,\Delta \varphi_D)$. The time-reversed trajectory is then $\textbf{b}_i=(\delta_D,\Delta \varphi_D, \dots , \delta_1,\Delta \varphi_1)$. Note that it is imperative that the order of $\delta$ and $\Delta \varphi$ does not change.}

Finally, the symmetric KL divergence (Jeffreys divergence) between the annihilation and creation is defined as half the sum of KLD(forward creation $||$ backward annihilation) + KLD(forward annihilation $||$ backward creation ).
See below for the estimation method.

\red{The results may depend on the choice of $D$. On the one hand, smaller $D$ implies more coarse graining, resulting in a looser lower bound. On the other hand, higher $D$ implies a higher dimension for the estimate of the divergence, usually leading to larger statistical errors. Since, for any $D$, the estimate provides a lower bound for the actual EP, We may optimize its value. We use $D=4$ for bacteria and $D=11$ for cells.}

\subsubsection*{Estimation of Kullback-Leibler divergence}

Let $F$ and $B$ denote continuous distributions over $\mathbb{R}^D$ such that $F$ is absolutely continuous with respect to $B$. The densities are denoted $f(x)$ and $b(x)$, respectively. We would like to estimate the Kullback-Leibler divergence of $F$ with respect to $B$, 
\begin{equation}
    D_{\rm KL} (F || B) = \int_{\mathbb{R}^D} f(x) \ln \frac{f(x)}{b(x)} d x ,
\end{equation}    
using independent samples $\textbf{f}_1 \dots \textbf{f}_N$ and $\textbf{b}_1 \dots \textbf{b}_M$, taken from $F$ and $B$ respectively.
Throughout, we take $0 \ln 0/0=0$.

Here, the divergence is estimated using the $k$-nearest neighbors ($k$NN) suggested by Wang et al \cite{wang2009divergence}. The main idea is as follows. We note that the $D_{\rm KL}$ can be written as an average, 
\begin{equation}
    D_{\rm KL} (F || B) = \left< \ln \frac{f(X)}{b(X)} \right> ,
    \label{eq:DKLasAv}
\end{equation}
where $X$ denotes a random variable with distribution $F$.
Let $k \ll N,M$.
Locally, around a sample point $\textbf{f}_i$, the density $f(x)$
can be approximated as
\begin{equation}
    f(\textbf{f}_i) \simeq \frac{k}{N-1} \frac{1}{V_D \rho_k^D (i)},
\end{equation}
where $V_D$ is the volume of the unit ball in $\mathbb{R}^D$ and $\rho_k (i)$ is the (Euclidean) distance from position $\textbf{f}_i$ of point $i$, to the $k$'s NN out of $\textbf{f}_1 \dots \textbf{f}_N$ (not counting $i$ itself, hence $N-1$).
Similarly, the density $b(x)$ around a sample point $i$, 
can be approximated as
\begin{equation}
    b(\textbf{f}_i) \simeq \frac{k}{M} \frac{1}{V_D \nu_k^D (i)},
\end{equation}
$\nu_k (i)$ is the distance from $\textbf{f}_i$ to the $k$'s NN out of $\textbf{b}_1 \dots \textbf{b}_M$.
Substituting into \eqref{eq:DKLasAv} yields,
\begin{equation}
    D_{\rm KL} (F || B) \simeq \frac{D}{N} \sum_{i=1}^N \frac{\nu_k(i)}{\rho_k(i)} + \ln \frac{M}{N-1} .
    \label{eq:DKLestimate}
\end{equation}
See \cite{wang2009divergence} for a proof \eqref{eq:DKLestimate} is a consistent estimator for $D_{\rm KL} (F || B)$.

To estimate the statistical error, we recall that the divergence is an average quantity.
Accordingly (see also \cite{beirlant1997nonparametric} for similar considerations for estimation of entropy),
we take the standard error to be
\begin{equation}
    \sigma_{\rm KL} (F || B) = \frac{1}{\sqrt{N}} {\rm std} \left[ \ln \frac{f(X)}{b(X)} \right] ,
    \label{eq:DKerr}
\end{equation}
where std denotes the standard deviation. Once again, approximating the densities using $k$NN yield an estimator for the standard error \cite{beirlant1997nonparametric}.

\red{The parameter $k$, setting the number of neighbors, should be taken to be proportional to $\sqrt{N}$~\cite{beirlant1997nonparametric}. Estimates reported used $k=\sqrt{N}$. 
}

\subsection*{\red{Simulating an active nemato-polar system}}

\red{We start from the standard description of an incompressible two-dimensional active nematic. The nematic orientation of the cells is described by the $\bm{Q}$ tensor which combines the cellular orientation $\bm{\hat{n}}$ and the local degree of nematic order $S\in[0,1]$ into a single tensorial order parameter $Q_{ij} = S(n_i n_j - \delta_{ij}/2)$. The motion of the cells are described by the velocity field $\bm{v}$ which abides by $\nabla.\bm{v}=0$. 

The dynamics of the nematic tensor are given by 
\begin{equation}
    D_t\bm{Q} = \lambda_Q S \bm{u} + \frac{1}{\gamma_Q}\bm{H}_Q + \bm{Q}.\bm{\omega} - \bm{\omega}.\bm{Q},
\end{equation}
where $D_t = \partial_t + \bm{v}.\nabla$ is the material derivative. $\bm{u}$ and $\bm{\omega}$ represent the strain rate and vorticity tensors, respectively. The molecular tensor is given by the functional derivative of the elastic free energy with respect to the nematic tensor, $\bm{H} = -\delta F/\delta \bm{Q}$ and $\gamma_Q$ is the rotational viscosity associated with the nematic field. The variable $\lambda_Q$ describes the tendency of the nematic field to align with the flow. 

To describe the polar orientation of the cells we introduce the vectorial order parameter $\bm{p}$ which describes the average local orientation of the self-propulsion direction of the cells. This follows similar dynamics to the nematic tensor, given by
\begin{equation}
    D_t\bm{p} = \lambda_p \bm{u}.\bm{p} + \frac{1}{\gamma_p}\bm{h}_p - \bm{\omega}.\bm{p}.
\end{equation}
Here, the molecular vector is given by the functional derivative of the elastic free energy with respect to the polar field, $\bm{h} = -\delta F/\delta \bm{p}$ and $\gamma_p$ is the rotational viscosity associated with the polar field. As before, the variable $\lambda_p$ describes the tendency of the polar field to align with the flow. 

The elastic free energy governs the relaxation dynamics of the $\bm{Q}$ and $\bm{p}$ fields. We split it into 3 parts,
\begin{equation}
    F = F_Q + F_C + F_p.
\end{equation}
The first part contains the nematic energy and describes the steric interactions between cells. It is given by 
\begin{equation}
    F_Q = \frac{1}{2}\int \Big[K_Q|\nabla\bm{Q}|^2 + \alpha_Q\textrm{tr}\bm{Q}^2(\textrm{tr}\bm{Q}^2 - 1)\Big] \textrm{d}A.
\end{equation}
In the expression above, $K_Q$ denotes the single elastic constant and $\alpha_Q$ controls the strength of the Landau terms which apply a soft constraint pushing the system into the nematic phase $S=1$.

The second part of the energy is the coupling term between the $\bm{Q}$ and $\bm{p}$ fields given by 
\begin{equation}
    F_C = -\frac{\alpha_C}{2}\bm{p}.\bm{Q}.\bm{p}.
\end{equation}
This energy is minimized when the $\bm{Q}$ and $\bm{p}$ fields align and is used to ensure they remain compatible descriptions of the cellular orientations.

The final part of the energy describes the tendency to locally align polarity and is given by 
\begin{equation}
    F_p = \frac{1}{2}\int \Big[K_p|\nabla\bm{p}|^2 + \frac{\alpha_p}{4}\bm{p}^4 - \frac{\beta_p}{2}\bm{p}^2\textrm{d}A.
\end{equation}
Here, $K_p$ is the elastic constant associated with gradients in the polar field and $\alpha_p$ and $\beta_p$ control the transition to polar order. This energy is employed predominantly to maintain the polar order parameter and keep the field continuous, thus it has a smaller magnitude than the other energies. We typically set $\beta_p=0$. This implies that the polar field becomes disordered when there is no nematic order, for example at the core of a defect. 

The dynamics of the velocity evolve according to 
\begin{equation}
    \label{eq:mom}
		\rho D_t \bm{v} = \eta \nabla^2\bm{v} -P\bm{I} + \nabla.\bm{\sigma}^p + \zeta_Q \nabla.\bm{Q} + \zeta_p\bm{p} - \mu\bm{v} .
\end{equation}
Here, $\rho$ is a constant density, $\eta$ is the shear viscosity, and $\mu$ is the friction coefficient. $\zeta_Q$ is the magnitude of the nematic active stress and $\zeta_p$ is the magnitude of the polar active force. $P$ if the pressure which is fixed by the incompressibility condition and $\bm{I}$ is the identity matrix.

Finally, the passive stress is given by 
\begin{equation}
    \bm{\sigma}^p = -\lambda_Q S\bm{H} + \bm{Q}.\bm{H} - \bm{H}.\bm{Q} -\lambda_{p}\bm{h}\odot\bm{p} - \bm{h}\wedge\bm{p} .
\end{equation}
The symmetric and antisymmetric tensor products are defined as
\begin{align}
    \bm{h}\odot\bm{p} &= \frac{1}{2} (h_ip_j + h_jp_i)\\
    \bm{h}\wedge\bm{p} &= \frac{1}{2}(h_ip_j - h_jp_i).
\end{align}

We simulate the dynamical equations on a $512\times512$ descretized grid using finite differences to approximate the gradients. We use a vorticity stream function approach to solve for incompressibility. We non-dimensionalize our equations using the length scale $\epsilon = \sqrt{K_Q/\alpha_Q} = 1$, and time scale $\tau = \gamma_Q/\alpha_Q = 1$, setting our simulation size to $L=128$. We measure energy in units of the elastic constant, $K_Q = 1$. The remaining parameters are given by $\rho = 1$, $\eta = 1$ and $\mu = 16$. We enforce low Reynolds number by neglecting the advective term in Eq.~\ref{eq:mom}. We simulate the system in the absence of flow alignment, $\lambda_Q = \lambda_p = 0$. To ensure the polar field follows the nematic field closely, we set $K_p = K_Q/100$, $\alpha_p = \alpha_Q/25$ and $\alpha_C = 4\alpha_Q/5$. Finally we fix the nematic active stress at $\zeta_Q = -0.2$ and the polar active force is taken from the range $\zeta_{P} \in [0.0, 0.0026125]$. 

We use this model to simulate specific creation and annihilation events, as this allows us to keep the grain boundary correlated with the orientation of the $+^1/_2$ defect. The system is initialized with zero flow, $\underline{v}=0$. To simulate defect creation, we initialize the director fields with a localized bend distortion, see  Extended Data Fig. 7a. If $\theta$ is the orientation of the polar field the bend distortion is given by
\begin{equation}
\theta(\underline{r}) = \frac{\pi}{2}+\exp\left(-\frac{r^2}{\sigma^2}\right)\sin\left(\frac{2\pi r_y}{L}\right)
\label{eq:in_cre}
\end{equation}
Here $\underline{r} = (r_x,r_y)$ is the position relative to the center of the simulation area of size $L$ and $\sigma = 11L/128$ gives the scale of the distortion. 
For defect annihilation, we initialize the system with a pair of defects situated either side of the center of the simulation area, see  Extended Data Fig. 7b. The field around a defect is given by
\begin{equation}
\theta(\underline{r}) = \frac{\phi_1 - \phi_2}{2}.
\label{eq:in_ann}
\end{equation}
Here $\phi_1$ and $\phi_2$ are the polar angles around the locations of the $+^1/_2$ and $-^1/_2$ defects, respectively. The defects are located at $\underline{r}_{\mp}=(\pm\Delta,0)$ and $\Delta$ is half the separation between the defects; we set $\Delta = 3L/32$. Finally, we reverse all polar vectors above the $x$ axis, which moves the grain boundary to the tail of the $+^1/_2$ defect.
}

\subsection*{\red{Field around a topological defect}}
\red{
As above we describe the nematic field by the tensor $\bm{Q} = S(\bm{n}\odot\bm{n} - \bm{I}/2)$ where $\bm{n}$ is a vector aligned with the nematic director. We now introduce the nematic orientation angle, $\psi$ which describes the average orientation of the nematic director at a point in space, hence 
\begin{equation}
    \hat{\bm{n}} = [\cos(\psi),\sin(\psi)].
\end{equation}
    
Topological defects within the nematic field are characterized by a half integer winding number $k=\pm ^1/_2$. The orientation of the nematic director field at a polar angle $\phi$ around a topological defect is given by:
\begin{equation}
    \psi = k\phi + \psi_0
\end{equation}
where we have introduced a phase $\psi_0$ which corresponds to a global rotation of the defect by an angle of $\frac{\psi_0}{1-k}$. Note, here we will adopt the convention that $\phi \in [-\pi,\pi]$ for convenience and $\psi_0=0$ which fixes the orientation of the defect relative to the frame of reference. Examples of these defects are given in Extended Data Fig. 3.

We describe the polar field with $\bm{p}$ and introduce the polar orientation angle, $\theta$, which is compatible with 
\begin{equation}
    \hat{\bm{p}} = p[\cos(\theta),\sin(\theta)].
\end{equation}
For the following calculations we will neglect variations in the order parameter, i.e. $S=p=1$ and assume perfect correlation between the polar and nematic fields, i.e. $\bm{p}.\bm{n} = \pm 1$.

A half integer winding number, $k=\pm^1/_2$, is incompatible with the polar field $\bm{p}$, and necessitates the introduction of a discontinuous jump in the polar director angle, $\theta$. This is the grain boundary over which $\bm{{p}}$ goes through a $\pi$ rotation which begins and ends at a topological defect. We introduce the grain boundary as follows
\begin{equation}
    \theta = k\phi + \psi_0 + \pi*(H(\phi - \phi_g) + m)
\end{equation}
$m\in\{0,1\}$ corresponds to a global rotation of $\hat{p}$ by $\pi$ and captures the additional symmetry of the polar field relative to a nematic field. The grain boundary is introduced with a Heaviside step function given by:
\begin{equation}
    H(\phi-\phi_g) = 
    \begin{cases}
        0, & \phi < \phi_g\\
        1, & \phi > \phi_g\\
    \end{cases}
\end{equation}
$\phi_g$ describes the polar angle at which the grain boundary approaches the core of the topological defect. Without loss of generality, we constrain $\phi_g\in (-\pi,\pi]$. The grain boundary and resulting polar field are shown in Extended Data Fig. 3.
}

\subsubsection*{\red{Flow field around a topological defect}}
\red{
At low Reynolds number, the flow caused by a force field $\bm{f}$ obeys the following equation
\begin{equation}
    \eta\nabla^2\bm{v} - \nabla P + \bm{f} = 0.
    \label{eq:stokes}
\end{equation}
Where $\eta$ is the viscosity and $P$ is the pressure. For an incompressible fluid, solutions to this equation can be found by convoluting the force field $\bm{f}$ with the two dimensional Oseen tensor $\bm{G}$, hence
\begin{equation}
    v_i(\bm{r}) = \int G_{ij}(\bm{r} - \bm{r}')f_j(\bm{r}') dA',
\end{equation}
where the Oseen tensor is given by
\begin{equation}
    G_{ij}(\bm{r}) = \frac{1}{4\pi\eta}\left[(\log{\frac{\mathcal{L}}{{r}}}-1)\delta_{ij} + \frac{r_ir_j}{r^2}\right]
\end{equation}

To calculate the flow around a defect, we assume that the active forces dominate, neglecting the passive stress tensor. In our model, the flow around a defect is dominated by the two active forces. The active force arising from the extensile active stress tensor is given by
\begin{equation}
    \bm{f}_Q = \zeta_Q\nabla.\bm{Q},
\end{equation}
and can be calculated as 
\begin{equation}
    \bm{f}_Q = \frac{\zeta_Q}{2r}\bm{\hat{x}}
\end{equation}
where $r$ is the distance from the defect. 

Following the calculation in \cite{giomi2014defect} one can arrive at the following flow field arising from the active nematic forces
\begin{equation}
    \bm{v}^Q(r,\phi) = \frac{\zeta_Q}{12\eta}\left[3(R-r) + r\cos(2\phi),r\sin(2\phi)\right].
\end{equation}
Due to the linearity of Eq.~\ref{eq:stokes}, we can simply add this to the additional velocity caused by the polar body force. 

The polar body force arising from self-propulsion is given by
\begin{equation}
    \bm{f}_p = \zeta_p\bm{p}.
\end{equation}
This can be calculated around a defect as
\begin{equation}
    \bm{f}_p = \zeta_p(-1)^m S(\phi , \phi_{g})\cos \left(\frac{\phi}{2} \right)\hat{\mathbf{x}} + \zeta_p(-1)^m S(\phi , \phi_{g})\sin \left(\frac{\phi}{2} \right)\hat{\mathbf{y}}.
\end{equation}
Here we have introduced the sign function defined by 
\begin{equation}
    \begin{aligned}
        S(\phi , \phi_{g}) = \begin{cases}
        +1 \quad & \phi \leq \phi_{g} \\
        -1 \quad & \phi > \phi_{g}
        \end{cases}
    \end{aligned}
\end{equation}

Following a similar process to that shown in Ref~\cite{giomi2014defect}, one can arrive at a flow field that results from the active polar body force, which we denote $\bm{v}^p$. The equation representing $\bm{v}^p$ is very long and a full derivation is given in the appendix, along with the equivalent calculation for a $-^1/_2$ defect. The final flow field around an active $+^1/_2$ defect is given by $\bm{v} = \bm{v}^Q + \bm{v}^p$ and is shown in Fig.~\ref{fig:F5}e.

To assess the active self-propulsion of a $+^1/_2$ defect, we evaluate the generated velocity at the defect core, i.e. at $r=0$. This gives 
\begin{equation}
    v(r=0) = \frac{R\zeta_Q}{4\pi\eta}\left[\pi + \frac{\zeta_p}{\zeta_Q}R\sin\left(\frac{\phi_G}{2}\right), -\frac{\zeta_p}{\zeta_Q}R\cos\left(\frac{\phi_G}{2}\right)\right].
    \label{eq:selfP}
\end{equation}
The argument of this velocity vector is displayed in Fig.~\ref{fig:F5}f (evaluated for $R=1$ and $\eta=1$).

The torque on the defect is estimated from the curl of the velocity field at $r=0$, which is calculated numerically and displayed in Fig.~\ref{fig:F5}g.
}

\subsubsection*{Total force on a topological defect core due to polar forcing}

Calculating the flow field around the defect is often a difficult process, however calculating the net force at the defect core is comparatively easy and gives very similar results. We include it here for completeness. 

We now calculate the total additional force on the defect core due to polar forcing by integrating the polar force azimuthally around the core of the defect. Without loss of generality we set $\psi_0 = 0$, which fixes the orientation of the defect. This results in the tail of the $+^1/_2$ defect, or one of the legs of the $-^1/_2$ defect, pointing in the $\phi=0$ direction, which we set as the $+\hat{\underline{x}}$ direction. 
	
The grain boundary breaks the mirror symmetry of a $+^1/_2$ defect if is introduced at any angle other than that opposite to the tail of the defect. Thus in most cases, the $+^1/_2$ defect becomes chiral. 
		
The total body force around the core of the defect with charge $k$ is given by. 
	
\begin{align}
    \underline{F}_p &= \oint \underline{f}_p(\phi) \textrm{d}\phi\\
    &\sim \zeta_p\oint \underline{\hat{p}}(\phi) \textrm{d}\phi\\
\end{align}
It is convenient to evaluate this integral piecewise to find the solution:
\begin{align*}
    \underline{F} &\sim \zeta_p\int_{-\pi}^{\phi_g} \underline{\hat{p}}(\phi) \textrm{d}\phi + \zeta_p\int_{\phi_g}^{\pi} \underline{\hat{p}}(\phi) \textrm{d}\phi \\
    & = \zeta_p\int_{-\pi}^{\phi_g} [\cos(k\phi + m\pi),\sin(k\phi + m\pi)] \textrm{d}\phi +\\
    & \zeta_p\int_{\phi_g}^{\pi} [\cos(k\phi + (m+1)\pi),\sin(k\phi + (m+1)\pi)] \textrm{d}\phi\\
    & = \zeta_p\left[\frac{[\sin(k\phi + m\pi),-\cos(k\phi + m\pi)]}{k}\right]_{-\pi}^{\phi_g} +\\
    & \zeta_p\left[\frac{[\sin(k\phi + (m+1)\pi),-\cos(k\phi + (m+1)\pi)]}{k}\right]_{\phi_g}^{\pi}\\
    & = \frac{2\zeta_p}{k}\left[\sin\left(k\phi_g + m\pi\right) ,-\cos\left(k\phi_g + m\pi\right)\right]
\end{align*}

This force is shown as the large red arrow in Extended Data Fig. 3.
Some important things to note here. 

First, there is a net translational force on both $\pm^1/_2$ defects, meaning that we expect both defects to self-propel. 

If $\phi_g = 0$, as is the case immediately after creation, the additional force is perfectly perpendicular to the symmetry axis of the defect. When combined with the force from the extensile stress, this would imply a self-propulsion at an angle to the tail for $+^1/_2$ defects. 

The additional force on the defects breaks chiral symmetry provided it does not align with a mirror symmetry axis of the defect. The mirror symmetry axes of the defect are given by $\phi_s = n\pi/2(k-1)$, $n\in\mathbb{Z}$. In addition, $m$ rotates the force by $\pi$ meaning it changes the sign of the additional force. This switches the chirality of the effect. As mentioned in the main text, $m$ arises from a spontaneously broken symmetry. 

\red{Finally, the force on the core of a $k=+^1/_2$ defect is perfectly aligned with the flow predicted by the Stokes flow calculation. If we add this to the net force arising from the extensile active stress, which is given by $\underline{F}_Q\sim[\zeta_Q,0]$ we arrive at 
\begin{equation}
    \underline{F} = \left[\zeta_Q + \zeta_p\sin\left(\phi_g/2\right),-\zeta_p\cos(\phi_g/2)\right],
    \label{eq:netF}
\end{equation}
which shows the same behavior and symmetries as Eq.~\ref{eq:selfP}, see Extended Data Fig. 3c. Thus the self-propelled motion of the $+^1/_2$ defects can be simply understood by considering the linear combination of the active forces at the core. }

\subsubsection*{Total torque on a topological defect core due to polar forcing}

We can assess the active torsional force around a defect caused by the polar active force by azimuthally integrating the azimuthal component of the polar force. This gives an equivalent picture to the predicted vorticity at the core of the defect. The curl of the polar force is given by:

\begin{equation}
    \tau = \oint \underline{f}.\underline{\textrm{d}l}
\end{equation}
This integration is performed on a circle around the defect and $\underline{dl}$ is tangent to that circle. Thus we can write:
\begin{equation}
    \tau \sim \zeta_p\int_{-\pi}^{\pi} \underline{\hat{p}}.[-\sin(\phi),\cos(\phi)] \textrm{d}\phi
\end{equation}
We can simplify this with angle addition formula to obtain:
\begin{align}
    \tau &\sim \zeta_p\oint_{-\pi}^{\pi}\sin(\theta - \phi) \textrm{d}\phi\\
    &= \zeta_p\oint_{-\pi}^{\pi} \sin((k-1)\phi + \pi*(H(\phi - \phi_g) + m)) \textrm{d}\phi\\
    &= \zeta_p\left[\frac{-\cos((k-1)\phi + m\pi)}{k-1}\right]_{-\pi}^{\phi_g} +\\
    & \zeta_p\left[\frac{-\cos((k-1)\phi + (m+1)\pi)}{k-1}\right]_{\phi_g}^{\pi}\\
    &= \frac{\zeta_p}{k-1}\left[2\cos((k-1)\phi_g + (m+1)\pi) + \right.\\
    & \quad \left. \cos(m\pi) - \cos((2k+m-1)\pi)\right]\\
    &= \frac{\zeta_p}{1-k}\cos((k-1)\phi_g + m\pi)
    \label{eq:tau}
\end{align}
This torque is shown as the large green arrow in Extended Data Fig. 3.

Some things to note here:

First, the sign of the torque depends on $m$, thus is selected by a spontaneous symmetry break. 

Second, the magnitude of the torque depends on the orientation of the grain boundary. The torque disappears when $(k-1)\phi_g + m\pi = (n+0.5)\pi$, $n\in\mathbb{Z}$. This is equivalent to when the net additional net force aligns with an axis of mirror symmetry. 

Third, the magnitude of the torque additionally depends on $k$, with $+^1/_2$ defects experiencing three times more torque on their core than $-^1/2$ defects. 

\red{Finally, since the curl of the force arising from the active stress is zero around the core of a defect, any net vorticity around the defect must be arise from the curl of the body force. We plot $\tau$ around a $+^1/_2$ defect as a function of $\phi_g$ and $\zeta_p$ in Extended Data Fig. 3d, and we see that it is a good predictor for the vorticity around a defect (Fig.~\ref{fig:F5}g.}


\bibliography{references}

@article{grossmann2020particle,
  title={A particle-field approach bridges phase separation and collective motion in active matter},
  author={Gro{\ss}mann, Robert and Aranson, Igor S and Peruani, Fernando},
  journal={Nature communications},
  volume={11},
  number={1},
  pages={5365},
  year={2020},
  publisher={Nature Publishing Group UK London}
}

@article{giomi2014defect,
  title={Defect dynamics in active nematics},
  author={Giomi, Luca and Bowick, Mark J and Mishra, Prashant and Sknepnek, Rastko and Cristina Marchetti, M},
  journal={Philosophical Transactions of the Royal Society A: Mathematical, Physical and Engineering Sciences},
  volume={372},
  number={2029},
  pages={20130365},
  year={2014},
  publisher={The Royal Society Publishing}
}

@article{pearce2021orientational,
  title={Orientational correlations in active and passive nematic defects},
  author={Pearce, DJG and Nambisan, J and Ellis, PW and Fernandez-Nieves, A and Giomi, L},
  journal={Physical Review Letters},
  volume={127},
  number={19},
  pages={197801},
  year={2021},
  publisher={APS}
}

@article{marchetti2013hydrodynamics,
  title={Hydrodynamics of soft active matter},
  author={Marchetti, M Cristina and Joanny, Jean-Fran{\c{c}}ois and Ramaswamy, Sriram and Liverpool, Tanniemola B and Prost, Jacques and Rao, Madan and Simha, R Aditi},
  journal={Reviews of modern physics},
  volume={85},
  number={3},
  pages={1143--1189},
  year={2013},
  publisher={APS}
}

@article{shimaya2022tilt,
  title={Tilt-induced polar order and topological defects in growing bacterial populations},
  author={Shimaya, Takuro and Takeuchi, Kazumasa A},
  journal={PNAS nexus},
  volume={1},
  number={5},
  pages={pgac269},
  year={2022},
  publisher={Oxford University Press}
}

@article{copenhagen2021topological,
  title={Topological defects promote layer formation in Myxococcus xanthus colonies},
  author={Copenhagen, Katherine and Alert, Ricard and Wingreen, Ned S and Shaevitz, Joshua W},
  journal={Nature Physics},
  volume={17},
  number={2},
  pages={211--215},
  year={2021},
  publisher={Nature Publishing Group UK London}
}

@article{li2019data,
  title={Data-driven quantitative modeling of bacterial active nematics},
  author={Li, He and Shi, Xia-qing and Huang, Mingji and Chen, Xiao and Xiao, Minfeng and Liu, Chenli and Chat{\'e}, Hugues and Zhang, HP},
  journal={Proceedings of the National Academy of Sciences},
  volume={116},
  number={3},
  pages={777--785},
  year={2019},
  publisher={National Academy of Sciences}
}

@article{genkin2017topological,
  title={Topological defects in a living nematic ensnare swimming bacteria},
  author={Genkin, Mikhail M and Sokolov, Andrey and Lavrentovich, Oleg D and Aranson, Igor S},
  journal={Physical Review X},
  volume={7},
  number={1},
  pages={011029},
  year={2017},
  publisher={APS}
}

@article{meacock2021bacteria,
  title={Bacteria solve the problem of crowding by moving slowly},
  author={Meacock, Oliver J and Doostmohammadi, Amin and Foster, Kevin R and Yeomans, Julia M and Durham, William M},
  journal={Nature Physics},
  volume={17},
  number={2},
  pages={205--210},
  year={2021},
  publisher={Nature Publishing Group UK London}
}

@article{yashunsky2024topological,
  title={Topological defects in multi-layered swarming bacteria},
  author={Yashunsky, Victor and Pearce, Daniel JG and Ariel, Gil and Be’er, Avraham},
  journal={Soft Matter},
  volume={20},
  number={21},
  pages={4237--4245},
  year={2024},
  publisher={Royal Society of Chemistry}
}

@article{blanch2018turbulent,
  title={Turbulent dynamics of epithelial cell cultures},
 author={Blanch-Mercader, Carles and Yashunsky, Victor and Garcia, Simon and Duclos, Guillaume and Giomi, Luca and Silberzan, Pascal},
  journal={Physical review letters},
  volume={120},
  number={20},
  pages={208101},
  year={2018},
  publisher={APS}
}

@article{head2024spontaneous,
  title={Spontaneous self-constraint in active nematic flows},
  author={Head, Louise C and Dor{\'e}, Claire and Keogh, Ryan R and Bonn, Lasse and Negro, Giuseppe and Marenduzzo, Davide and Doostmohammadi, Amin and Thijssen, Kristian and L{\'o}pez-Le{\'o}n, Teresa and Shendruk, Tyler N},
  journal={Nature Physics},
  volume={20},
  number={3},
  pages={492--500},
  year={2024},
  publisher={Nature Publishing Group UK London}
}

@article{pearce2021properties,
  title={Properties of twisted topological defects in 2D nematic liquid crystals},
  author={Pearce, Daniel JG and Kruse, Karsten},
  journal={Soft Matter},
  volume={17},
  number={31},
  pages={7408--7417},
  year={2021},
  publisher={Royal Society of Chemistry}
}

@article{giomi2015geometry,
  title={Geometry and topology of turbulence in active nematics},
  author={Giomi, Luca},
  journal={Physical Review X},
  volume={5},
  number={3},
  pages={031003},
  year={2015},
  publisher={APS}
}

@article{tang2017orientation,
  title={Orientation of topological defects in 2D nematic liquid crystals},
  author={Tang, Xingzhou and Selinger, Jonathan V},
  journal={Soft matter},
  volume={13},
  number={32},
  pages={5481--5490},
  year={2017},
  publisher={Royal Society of Chemistry}
}

@article{shankar2018defect,
  title={Defect unbinding in active nematics},
  author={Shankar, Suraj and Ramaswamy, Sriram and Marchetti, M Cristina and Bowick, Mark J},
  journal={Physical review letters},
  volume={121},
  number={10},
  pages={108002},
  year={2018},
  publisher={APS}
}

@article{vromans2016orientational,
  title={Orientational properties of nematic disclinations},
  author={Vromans, Arthur J and Giomi, Luca},
  journal={Soft matter},
  volume={12},
  number={30},
  pages={6490--6495},
  year={2016},
  publisher={Royal Society of Chemistry}
}

@article{seifert2012stochastic,
  title={Stochastic thermodynamics, fluctuation theorems and molecular machines},
  author={Seifert, Udo},
  journal={Reports on progress in physics},
  volume={75},
  number={12},
  pages={126001},
  year={2012},
  publisher={IOP Publishing}
}

@article{ro2022model,
  title={Model-free measurement of local entropy production and extractable work in active matter},
  author={Ro, Sunghan and Guo, Buming and Shih, Aaron and Phan, Trung V and Austin, Robert H and Levine, Dov and Chaikin, Paul M and Martiniani, Stefano},
  journal={Physical review letters},
  volume={129},
  number={22},
  pages={220601},
  year={2022},
  publisher={APS}
}

@article{fodor2022irreversibility,
  title={Irreversibility and biased ensembles in active matter: Insights from stochastic thermodynamics},
  author={Fodor, {\'E}tienne and Jack, Robert L and Cates, Michael E},
  journal={Annual Review of Condensed Matter Physics},
  volume={13},
  number={1},
  pages={215--238},
  year={2022},
  publisher={Annual Reviews}
}

@article{be2019statistical,
  title={A statistical physics view of swarming bacteria},
  author={Be’er, Avraham and Ariel, Gil},
  journal={Movement ecology},
  volume={7},
  pages={1--17},
  year={2019},
  publisher={Springer}
}

@article{aranson2022bacterial,
  title={Bacterial active matter},
  author={Aranson, Igor S},
  journal={Reports on Progress in Physics},
  volume={85},
  number={7},
  pages={076601},
  year={2022},
  publisher={IOP Publishing}
}

@article{sokolov2010swimming,
  title={Swimming bacteria power microscopic gears},
  author={Sokolov, Andrey and Apodaca, Mario M and Grzybowski, Bartosz A and Aranson, Igor S},
  journal={Proceedings of the National Academy of Sciences},
  volume={107},
  number={3},
  pages={969--974},
  year={2010},
  publisher={National Academy of Sciences}
}

@article{yashunsky2022chiral,
  title={Chiral edge current in nematic cell monolayers},
  author={Yashunsky, V and Pearce, DJG and Blanch-Mercader, C and Ascione, F and Silberzan, Pascal and Giomi, L},
  journal={Physical Review X},
  volume={12},
  number={4},
  pages={041017},
  year={2022},
  publisher={APS}
}

@article{serra2023defect,
  title={Defect-mediated dynamics of coherent structures in active nematics},
  author={Serra, Mattia and Lemma, Linnea and Giomi, Luca and Dogic, Zvonimir and Mahadevan, Lakshminarayanan},
  journal={Nature Physics},
  volume={19},
  number={9},
  pages={1355--1361},
  year={2023},
  publisher={Nature Publishing Group UK London}
}

@article{zhou2014living,
  title={Living liquid crystals},
  author={Zhou, Shuang and Sokolov, Andrey and Lavrentovich, Oleg D and Aranson, Igor S},
  journal={Biophysical Journal},
  volume={106},
  number={2},
  pages={420a},
  year={2014},
  publisher={Elsevier}
}

@article{guillamat2017taming,
  title={Taming active turbulence with patterned soft interfaces},
  author={Guillamat Bassedas, Pau and Ign{\'e}s i Mullol, Jordi and Sagu{\'e}s i Mestre, Francesc},
  journal={Nature Communications, 2017, vol. 8, num. 564},
  year={2017},
  publisher={Nature Publishing Group}
}

@article{sanchez2012spontaneous,
  title={Spontaneous motion in hierarchically assembled active matter},
  author={Sanchez, Tim and Chen, Daniel TN and DeCamp, Stephen J and Heymann, Michael and Dogic, Zvonimir},
  journal={Nature},
  volume={491},
  number={7424},
  pages={431--434},
  year={2012},
  publisher={Nature Publishing Group UK London}
}

@article{guillamat2018active,
  title={Active nematic emulsions},
  author={Guillamat, Pau and Kos, {\v{Z}}iga and Hardo{\"u}in, J{\'e}r{\^o}me and Ign{\'e}s-Mullol, Jordi and Ravnik, Miha and Sagu{\'e}s, Francesc},
  journal={Science advances},
  volume={4},
  number={4},
  pages={eaao1470},
  year={2018},
  publisher={American Association for the Advancement of Science}
}

@article{tan2019topological,
  title={Topological chaos in active nematics},
  author={Tan, Amanda J and Roberts, Eric and Smith, Spencer A and Olvera, Ulyses Alvarado and Arteaga, Jorge and Fortini, Sam and Mitchell, Kevin A and Hirst, Linda S},
  journal={Nature Physics},
  volume={15},
  number={10},
  pages={1033--1039},
  year={2019},
  publisher={Nature Publishing Group UK London}
}

@article{kumar2018tunable,
  title={Tunable structure and dynamics of active liquid crystals},
  author={Kumar, Nitin and Zhang, Rui and De Pablo, Juan J and Gardel, Margaret L},
  journal={Science advances},
  volume={4},
  number={10},
  pages={eaat7779},
  year={2018},
  publisher={American Association for the Advancement of Science}
}

@article{de2025hidden,
  title={Hidden order in active nematic defects},
  author={de la Cotte, Alexis and Pearce, Daniel JG and Nambisan, Jyothishraj and Puggioni, Leonardo and Levy, Aviv and Giomi, Luca and Fernandez-Nieves, Alberto},
  journal={Proceedings of the National Academy of Sciences},
  volume={122},
  number={40},
  pages={e2512147122},
  year={2025},
  publisher={National Academy of Sciences}
}

@article{vaidvziulyte2022persistent,
  title={Persistent cell migration emerges from a coupling between protrusion dynamics and polarized trafficking},
  author={Vaid{\v{z}}iulyt{\.e}, Kotryna and Mac{\'e}, Anne-Sophie and Battistella, Aude and Beng, William and Schauer, Kristine and Coppey, Mathieu},
  journal={Elife},
  volume={11},
  pages={e69229},
  year={2022},
  publisher={eLife Sciences Publications Limited}
}

@article{doostmohammadi2022physics,
  title={Physics of liquid crystals in cell biology},
  author={Doostmohammadi, Amin and Ladoux, Benoit},
  journal={Trends in cell biology},
  volume={32},
  number={2},
  pages={140--150},
  year={2022},
  publisher={Elsevier}
}

@article{beta2023actin,
  title={From actin waves to mechanism and back: How theory aids biological understanding},
  author={Beta, Carsten and Edelstein-Keshet, Leah and Gov, Nir and Yochelis, Arik},
  journal={Elife},
  volume={12},
  pages={e87181},
  year={2023},
  publisher={eLife Sciences Publications, Ltd}
}

@article{balasubramaniam2021investigating,
  title={Investigating the nature of active forces in tissues reveals how contractile cells can form extensile monolayers},
  author={Balasubramaniam, Lakshmi and Doostmohammadi, Amin and Saw, Thuan Beng and Narayana, Gautham Hari Narayana Sankara and Mueller, Romain and Dang, Tien and Thomas, Minnah and Gupta, Shafali and Sonam, Surabhi and Yap, Alpha S and others},
  journal={Nature materials},
  volume={20},
  number={8},
  pages={1156--1166},
  year={2021},
  publisher={Nature Publishing Group UK London}
}

@article{ienaga2023geometric,
  title={Geometric confinement guides topological defect pairings and emergent flow in nematic cell populations},
  author={Ienaga, Ryo and Beppu, Kazusa and Maeda, Yusuke T},
  journal={Soft Matter},
  volume={19},
  number={26},
  pages={5016--5028},
  year={2023},
  publisher={Royal Society of Chemistry}
}

@article{duclos2017topological,
  title={Topological defects in confined populations of spindle-shaped cells},
  author={Duclos, Guillaume and Erlenk{\"a}mper, Christoph and Joanny, Jean-Fran{\c{c}}ois and Silberzan, Pascal},
  journal={Nature Physics},
  volume={13},
  number={1},
  pages={58--62},
  year={2017},
  publisher={Nature Publishing Group UK London}
}

@article{kawaguchi2017topological,
  title={Topological defects control collective dynamics in neural progenitor cell cultures},
  author={Kawaguchi, Kyogo and Kageyama, Ryoichiro and Sano, Masaki},
  journal={Nature},
  volume={545},
  number={7654},
  pages={327--331},
  year={2017},
  publisher={Nature Publishing Group UK London}
}

@article{shankar2022topological,
  title={Topological active matter},
  author={Shankar, Suraj and Souslov, Anton and Bowick, Mark J and Marchetti, M Cristina and Vitelli, Vincenzo},
  journal={Nature Reviews Physics},
  volume={4},
  number={6},
  pages={380--398},
  year={2022},
  publisher={Nature Publishing Group UK London}
}

@article{saw2017topological,
  title={Topological defects in epithelia govern cell death and extrusion},
  author={Saw, Thuan Beng and Doostmohammadi, Amin and Nier, Vincent and Kocgozlu, Leyla and Thampi, Sumesh and Toyama, Yusuke and Marcq, Philippe and Lim, Chwee Teck and Yeomans, Julia M and Ladoux, Benoit},
  journal={Nature},
  volume={544},
  number={7649},
  pages={212--216},
  year={2017},
  publisher={Nature Publishing Group UK London}
}

@article{ardavseva2025beyond,
  title={Beyond Dipolar Activity: Quadrupolar Stress Drives Collapse of Nematic Order on Frictional Substrates},
  author={Arda{\v{s}}eva, Aleksandra and V{\'e}lez-Cer{\'o}n, Ignasi and Pedersen, Martin Cramer and Ign{\'e}s-Mullol, Jordi and Sagu{\'e}s, Francesc and Doostmohammadi, Amin},
  journal={Physical Review Letters},
  volume={134},
  number={8},
  pages={088301},
  year={2025},
  publisher={APS}
}

@article{patelli2019understanding,
  title={Understanding dense active nematics from microscopic models},
  author={Patelli, Aurelio and Djafer-Cherif, Ilyas and Aranson, Igor S and Bertin, Eric and Chat{\'e}, Hugues},
  journal={Physical review letters},
  volume={123},
  number={25},
  pages={258001},
  year={2019},
  publisher={APS}
}

@article{thampi2014vorticity,
  title={Vorticity, defects and correlations in active turbulence},
  author={Thampi, Sumesh P and Golestanian, Ramin and Yeomans, Julia M},
  journal={Philosophical Transactions of the Royal Society A: Mathematical, Physical and Engineering Sciences},
  volume={372},
  number={2029},
  pages={20130366},
  year={2014},
  publisher={The Royal Society Publishing}
}

@article{bonn2022fluctuation,
  title={Fluctuation-induced dynamics of nematic topological defects},
  author={Bonn, Lasse and Arda{\v{s}}eva, Aleksandra and Mueller, Romain and Shendruk, Tyler N and Doostmohammadi, Amin},
  journal={Physical Review E},
  volume={106},
  number={4},
  pages={044706},
  year={2022},
  publisher={APS}
}

@article{wang2009divergence,
  title={Divergence estimation for multidimensional densities via $ k $-Nearest-Neighbor distances},
  author={Wang, Qing and Kulkarni, Sanjeev R and Verd{\'u}, Sergio},
  journal={IEEE Transactions on Information Theory},
  volume={55},
  number={5},
  pages={2392--2405},
  year={2009},
  publisher={IEEE}
}

@article{beirlant1997nonparametric,
  title={Nonparametric entropy estimation: An overview},
  author={Beirlant, Jan and Dudewicz, Edward J and Gy{\"o}rfi, L{\'a}szl{\'o} and Van der Meulen, Edward C and others},
  journal={International Journal of Mathematical and Statistical Sciences},
  volume={6},
  number={1},
  pages={17--39},
  year={1997},
  publisher={THESAURUS PUBLISHING}
}

@article{martinez2021scaling,
  title={Scaling regimes of active turbulence with external dissipation},
  author={Mart{\'\i}nez-Prat, Berta and Alert, Ricard and Meng, Fanlong and Ign{\'e}s-Mullol, Jordi and Joanny, Jean-Fran{\c{c}}ois and Casademunt, Jaume and Golestanian, Ramin and Sagu{\'e}s, Francesc},
  journal={Physical Review X},
  volume={11},
  number={3},
  pages={031065},
  year={2021},
  publisher={APS}
}

@article{venkatesh2025interplay,
  title={The Interplay of Polar and Nematic Order in Active Matter: Implications for Non-Equilibrium Physics and Biology},
  author={Venkatesh, Varun and de Graaf Sousa, Niels and Doostmohammadi, Amin},
  journal={Journal of Physics A: Mathematical and Theoretical},
  year={2025}
}

@article{huber2018emergence,
  title={Emergence of coexisting ordered states in active matter systems},
  author={Huber, L and Suzuki, R and Kr{\"u}ger, T and Frey, E and Bausch, AR},
  journal={Science},
  volume={361},
  number={6399},
  pages={255--258},
  year={2018},
  publisher={American Association for the Advancement of Science}
}

@article{shi2018self,
  title={Self-propelled rods: Linking alignment-dominated and repulsion-dominated active matter},
  author={Shi, Xia-qing and Chat{\'e}, Hugues},
  journal={arXiv preprint arXiv:1807.00294},
  year={2018}
}

@article{amiri2022unifying,
  title={Unifying polar and nematic active matter: emergence and co-existence of half-integer and full-integer topological defects},
  author={Amiri, Aboutaleb and Mueller, Romain and Doostmohammadi, Amin},
  journal={Journal of Physics A: Mathematical and Theoretical},
  volume={55},
  number={9},
  pages={094002},
  year={2022},
  publisher={IOP Publishing}
}

@article{lacroix2024emergence,
  title={Emergence of bidirectional cell laning from collective contact guidance},
  author={Lacroix, Mathilde and Smeets, Bart and Blanch-Mercader, Carles and Bell, Samuel and Giuglaris, Caroline and Chen, Hsiang-Ying and Prost, Jacques and Silberzan, Pascal},
  journal={Nature Physics},
  volume={20},
  number={8},
  pages={1324--1331},
  year={2024},
  publisher={Nature Publishing Group UK London}
}

@article{maroudas2021topological,
  title={Topological defects in the nematic order of actin fibres as organization centres of Hydra morphogenesis},
  author={Maroudas-Sacks, Yonit and Garion, Liora and Shani-Zerbib, Lital and Livshits, Anton and Braun, Erez and Keren, Kinneret},
  journal={Nature Physics},
  volume={17},
  number={2},
  pages={251--259},
  year={2021},
  publisher={Nature Publishing Group UK London}
}

@article{han2025local,
  title={Local polar order controls mechanical stress and triggers layer formation in Myxococcus xanthus colonies},
  author={Han, Endao and Fei, Chenyi and Alert, Ricard and Copenhagen, Katherine and Koch, Matthias D and Wingreen, Ned S and Shaevitz, Joshua W},
  journal={Nature communications},
  volume={16},
  number={1},
  pages={952},
  year={2025},
  publisher={Nature Publishing Group UK London}
}

@article{de2025self,
  title={Self-propulsive active nematics},
  author={de Graaf Sousa, Niels and Andersen, Simon Guldager and Arda{\v{s}}eva, Aleksandra and Doostmohammadi, Amin},
  journal={Philosophical Transactions A},
  volume={383},
  number={2304},
  pages={20240272},
  year={2025},
  publisher={The Royal Society}
}

@article{vafa2025phase,
  title={Phase diagram, confining strings, and a new universality class in nematopolar matter},
  author={Vafa, Farzan and Doostmohammadi, Amin},
  journal={Europhysics Letters},
  year={2025}
}

@article{crocker1996methods,
  title={Methods of digital video microscopy for colloidal studies},
  author={Crocker, John C and Grier, David G},
  journal={Journal of colloid and interface science},
  volume={179},
  number={1},
  pages={298--310},
  year={1996},
  publisher={Elsevier}
}

@inproceedings{farneback2003two,
  title={Two-frame motion estimation based on polynomial expansion},
  author={Farneb{\"a}ck, Gunnar},
  booktitle={Scandinavian conference on Image analysis},
  pages={363--370},
  year={2003},
  organization={Springer}
}

@article{radhakrishnan2025irreversibility,
  title={Irreversibility in scalar active turbulence: The role of topological defects},
  author={Radhakrishnan, Byjesh N and Serafin, Francesco and Schmidt, Thomas L and Fodor, {\'E}tienne},
  journal={arXiv preprint arXiv:2507.06073},
  year={2025}
}

@article{doostmohammadi2016stabilization,
  title={Stabilization of active matter by flow-vortex lattices and defect ordering},
  author={Doostmohammadi, Amin and Adamer, Michael F and Thampi, Sumesh P and Yeomans, Julia M},
  journal={Nature communications},
  volume={7},
  number={1},
  pages={10557},
  year={2016},
  publisher={Nature Publishing Group UK London}
}

@article{lo2024spontaneous,
  title={Spontaneous rotations in epithelia as an interplay between cell polarity and boundaries},
  author={Lo Vecchio, S and Pertz, Olivier and Szopos, Marcela and Navoret, Laurent and Riveline, Daniel},
  journal={Nature Physics},
  volume={20},
  number={2},
  pages={322--331},
  year={2024},
  publisher={Nature Publishing Group UK London}
}
\bibliographystyle{refs.bst}

\newpage
\section*{Funding}
The research was supported by The Israel Science Foundation (grants no. 838/23, 2044/23, 382/23 and 161/24). DJGP and DC acknowledge funding from the Swiss National Science Foundation under starting grant TMSGI2 211367. This work was funded by the Deutsche Forschungsgemeinschaft (DFG, German Research Foundation) under grant number 396653815.

\section*{Author contributions}
A.B., G.A. and V.Y. designed research, A.B. and V.Y. performed the experiments, D.C., Y.A. and D.J.G.P. did the theoretical part, A.B., E.D.N., D.C., D.J.G.P., G.A. and V.Y. analysed data, and A.B., E.D.N., D.C., D.J.G.P., G.A. and V.Y. wrote the paper.

\section*{Competing interests}
There are no competing interests to declare.

\section*{Data and materials availability}
Original data will be available by the authors upon reasonable request. The code for the analyses of the experimental data can be accessed via GitLab at https://github.com/viciya/nematics. 


\clearpage

\begin{center}
\textbf{\Large Extended data}

\begin{figure*}
	\centering 
	\includegraphics[width=0.9\textwidth]{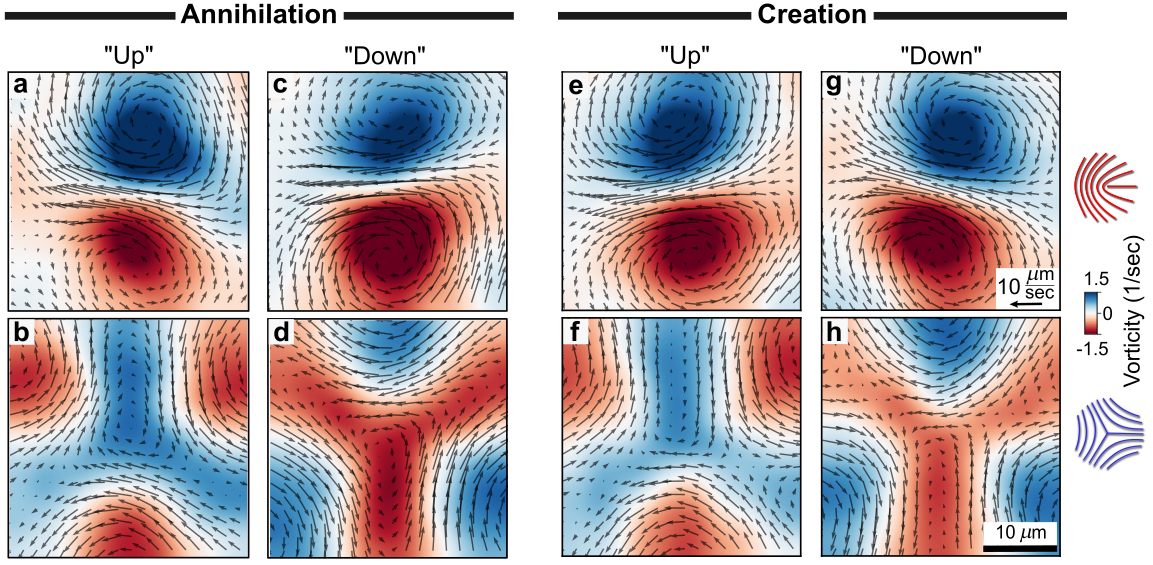}
    \captionsetup{justification=raggedright}
	\caption*{\textbf{Extended Data Fig. 1: Ensemble-averaged flow fields near $\pm^1/_2$ defect-pairs in bacterial swarm.}
    Average velocity fields (black arrows) and vorticity distributions (heatmap) around defects. Flow fields around paired $+^1/_2$ and $-^1/_2$ defects, averaged separately for the “up” (a, b, e, f) and “down” (c, d, g, h) configurations for annihilation (a, b, c, d) and creation events (e, f, g, h).}
	\label{fig:S1} 
\end{figure*}

\begin{figure*}
	\centering 
	\includegraphics[width=0.9\textwidth]{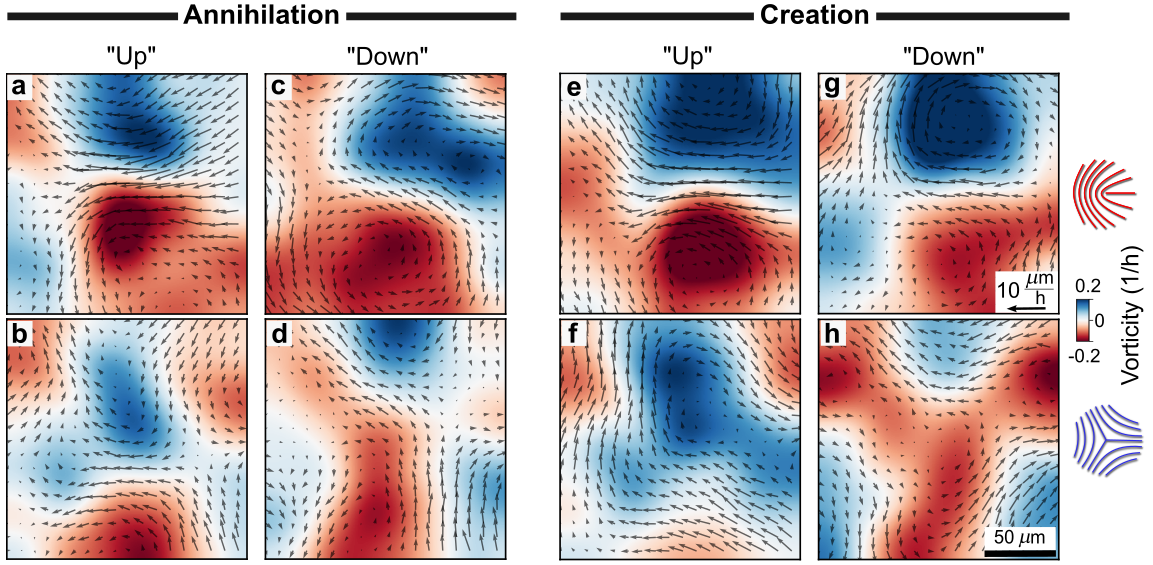}        \captionsetup{justification=justified, singlelinecheck=false}
	\caption*{\textbf{Extended Data Fig. 2: Ensemble-averaged flow fields near $\pm^1/_2$ defect-pairs in HBEC cell culture.}
    Average velocity fields (black arrows) and vorticity distributions (heatmap) around defects. Flow fields around paired $+^1/_2$ and $-^1/_2$ defects, averaged separately for the “up” (a, b, e, f) and “down” (c, d, g, h) configurations for annihilation (a, b, c, d) and creation events (e, f, g, h).}
	\label{fig:S2} 
\end{figure*}

\begin{figure*}
	\centering 
	\includegraphics[width=0.7\textwidth]{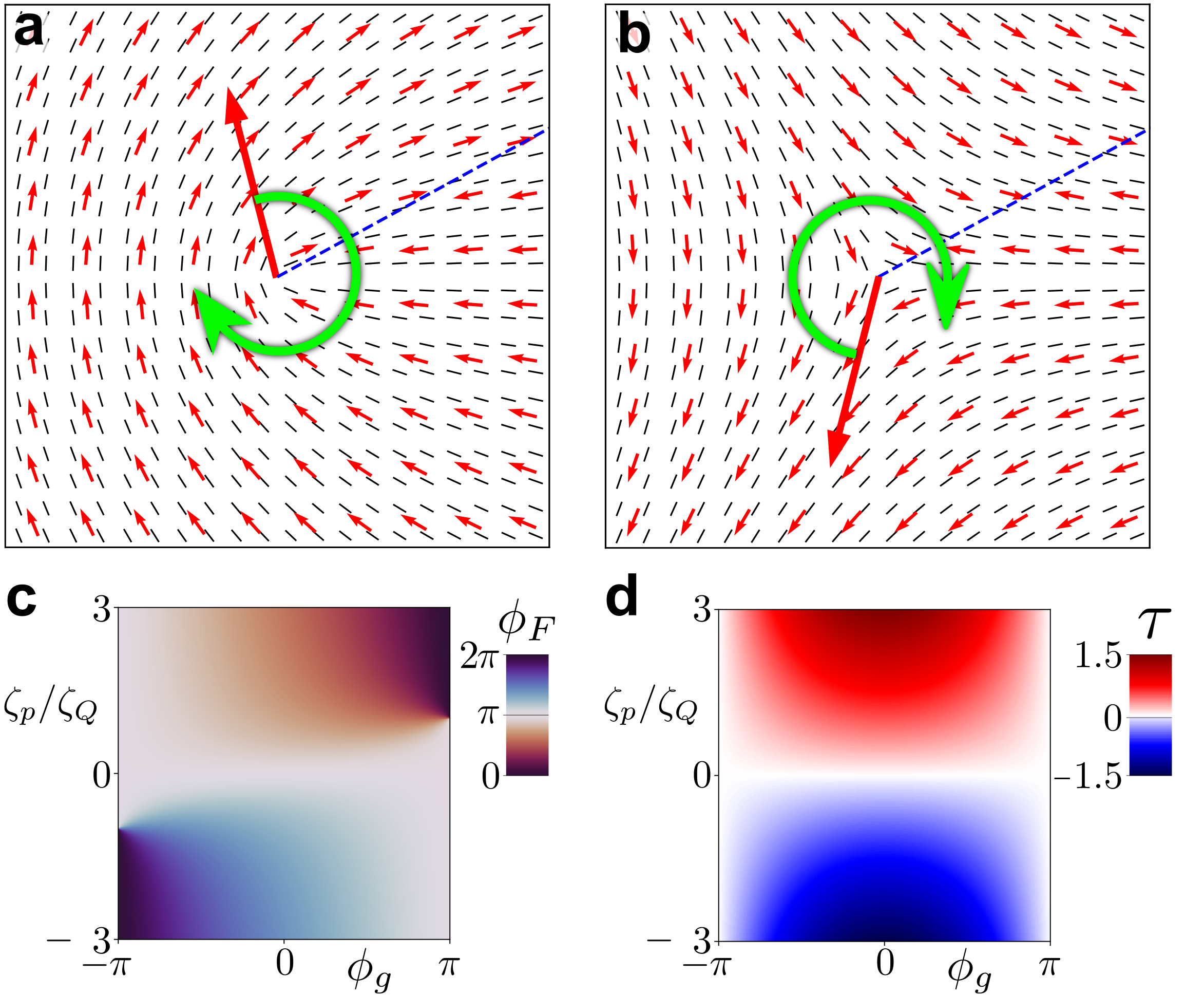}        \captionsetup{justification=justified, singlelinecheck=false}
	\caption*{\red{\textbf{Extended Data Fig. 3: Polar fields around half integer defects} (a) $+^1/_2$ and (b) $-^1/_2$ defects are shown in the nematic field (black lines). The polar field (small red arrows) around a half integer defect necessitates the existence of a grain boundary (blue dashed line at angle $\phi_g$) over which the polar field switches direction. At the core of the defect, the polar field has a net orientation (large red arrow) and net curl (green arrow). This corresponds to an additional force and torque on the defect core, respectively. The left shows a $k=0.5$ charge defect and the right shows a $k=-0.5$ defect. Both defects are presented with zero phase, $\psi_0=0$. The grain boundary in both cases is at $\phi_g = 0.5$ and $m=1$. The (c) orientation of the net force (Eq.~\ref{eq:netF}) and (d) curl of the polar body force (Eq.~\ref{eq:tau}) around the core of a $+^1/_2$ defect as a function of the postiion of the grain boundary, $\phi_g$, and the relative magnitude of the polar body force, $\zeta_p/\zeta_Q$. This simplified calculation shows the origin of the various symmetries of the Stokes flow around a defect shown in Fig.~\ref{fig:F5}f\&g.
    }}
	\label{fig:S3} 
\end{figure*}

\begin{figure*}[h] 
	\centering
    
	\includegraphics[width=0.45\textwidth]{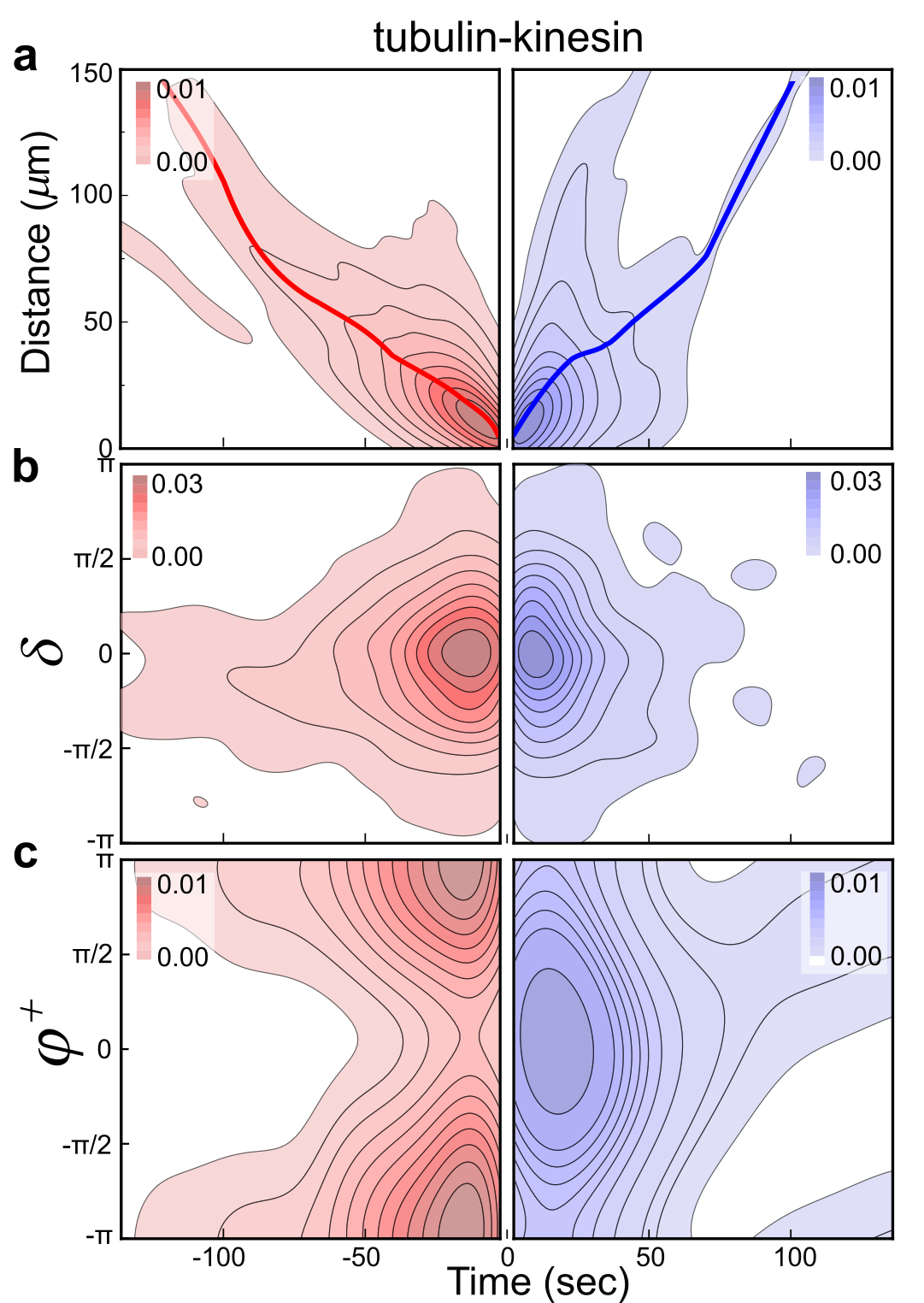}        \captionsetup{justification=justified, singlelinecheck=false}
	\caption*{\textbf{Extended Data Fig. 4:  Active nematic tubulin-kinesin fluid.} 
    \red{
    \textbf{(a)} Distance between $\pm ^1/_2$ defect pairs as a function of time after creation (blue) and before annihilation (red) in bacterial and cellular systems. The creation/annihilation instant is indicated by $t = 0$, with negative values corresponding to the time before annihilation and positive values to the time after creation. Solid lines represent the average trajectories for 29 annihilation and 31 creation events. Red and blue color intensities indicate the probability density.
    \textbf{(b)} Distribution of the pair orientation offset angle, $\delta = \varphi^+ - 3\varphi^- - \pi$, as a function of time for creation (blue) and annihilation (red) events. A value of $\delta = 0$ indicates synchronization of the nematic director fields between paired defects.    
    \textbf{(c)} Distribution of $\varphi^+$ as a function of time for creation (blue) and annihilation (red) events. The creation/annihilation instant is indicated by $t = 0$, with negative values indicating time before annihilation and positive values time after creation. Red and blue color intensities indicate the probability density. Defect dynamics data were extracted from Tan et al.~\cite{tan2019topological}.
    }
    }
	\label{fig:S4}
\end{figure*}

\begin{figure*}[h] 
	\centering
    \begin{minipage}{0.9\textwidth}
	\includegraphics[width=.75\textwidth]{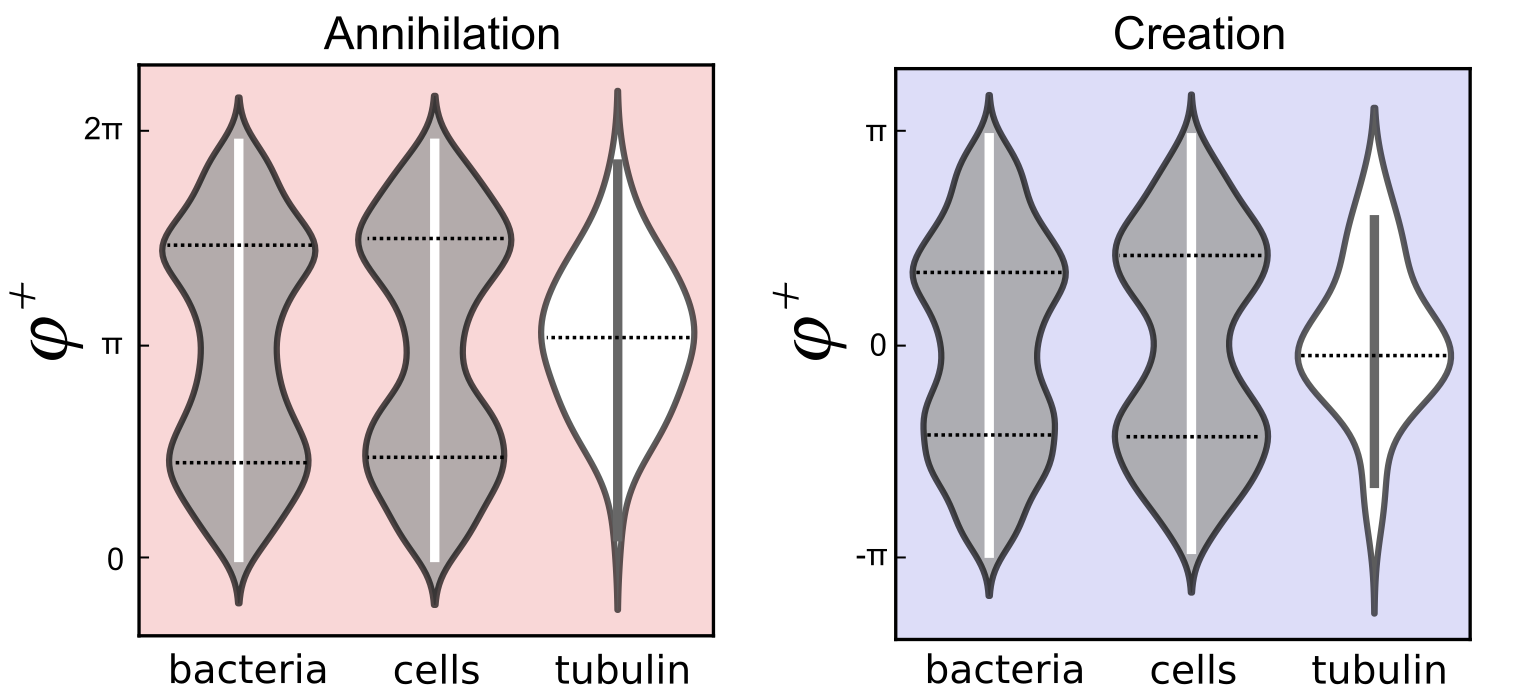}        
    \captionsetup{justification=justified, singlelinecheck=false}
	\caption*{\textbf{Extended Data Fig. 5:} \red{Configurations of $+^1/_2$ defects ($\varphi^+$) during annihilation and creation events, and their angular distributions, in bacterial, cellular, and tubulin–kinesin systems \cite{tan2019topological}. 
    }
    }
	\label{fig:S4}
    \end{minipage}
\end{figure*}

\begin{figure*}[h]
    \centering
    \includegraphics[width=0.85\textwidth]{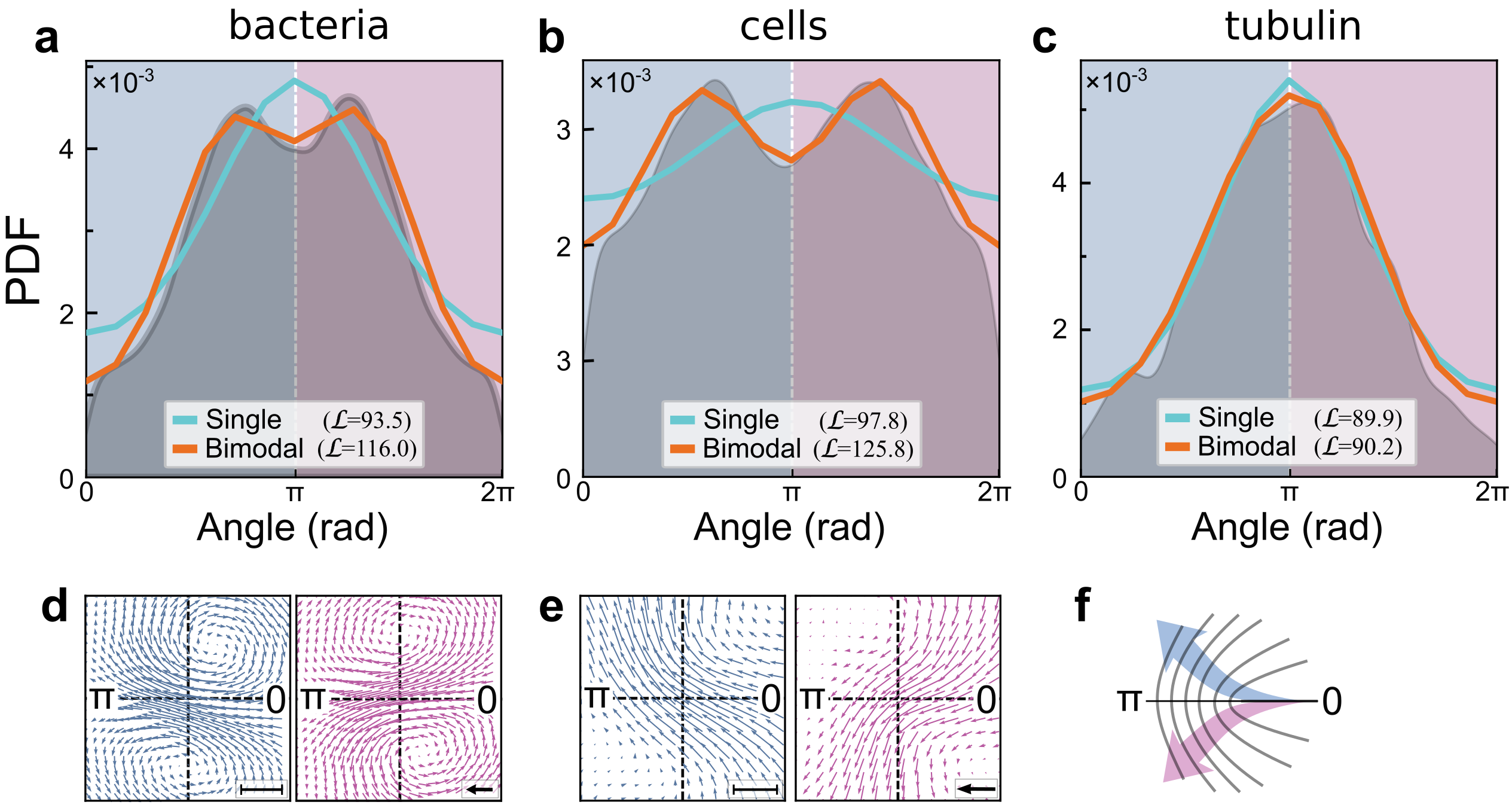}
    
    
    \captionsetup{justification=justified, singlelinecheck=false}
    \caption*{\textbf{Extended Data Fig. 6: Motion of unpaired $+^1/_2$ defects.} 
    \red{
    \textbf{(a–c)}, Angular probability distribution (gray area) of the $+^1/_2$ defect displacement angle in (a) bacteria, (b) cells, and (c) the tubulin–kinesin system \cite{tan2019topological}. Displacement angle is defined relative to the defect's symmetry plane, as illustrated in panel (f). The gray area represents the raw probability distribution. Superimposed curves show fits using a single von Mises distribution (cyan) and a bimodal von Mises distribution (orange). The log-likelihood ($\mathcal{L}$) for each fit is evaluated using the maximum likelihood estimate of the variance $\hat{\sigma}^2$ from the residuals ($\epsilon$):
    $\mathcal{L} = -\frac{N}{2} \left( \log(2\pi\hat{\sigma}^2) + 1 \right)$ where $N$ is the number of observations and $\hat{\sigma}^2 = \frac{1}{N} \sum \epsilon_i^2$ is the residual variance.    
    \textbf{(d, e)}, Average flow fields surrounding $+^1/_2$ defects. Flow fields are calculated by separately averaging defect trajectories based on their overall tilt angle ($\langle \phi_V \rangle$). Flows are shown for defects with an overall displacement to the right (blue-gray) and defects with an overall displacement to the left (pink). For the angular range $[0, 2\pi]$, left displacement corresponds to $\langle \phi_V \rangle > \pi$ and right displacement to $\langle \phi_V \rangle < \pi$. 
    Scale bars for bacteria and cells are $10~\mu \mathrm{m}$ and $50~\mu \mathrm{m}$, respectively. The scale of velocity magnitude indicated by black arrows are $10~\mu \mathrm{m}/\mathrm{sec}$ for bacteria and $10~\mu \mathrm{m}/\mathrm{h}$ for cells.
    
    Correlation between the average displacement angles $\langle \phi_V \rangle$ measured in the first and second halves of the trajectories (Spearman’s $\rho = 0.09$ for bacteria, $N = 3317$, and $0.07$ for cells, $N = 2557$; $p = 7 \times 10^{-8}$ for bacteria and $6 \times 10^{-4}$ for cells) indicates a weak but statistically significant directional persistence of the off-axis motion of $+^1/_2$ defects from creation to annihilation.
    
    }}
    \label{fig:S6}
\end{figure*}

\begin{figure*}[h]
    \centering
    \includegraphics[width=0.65\textwidth]{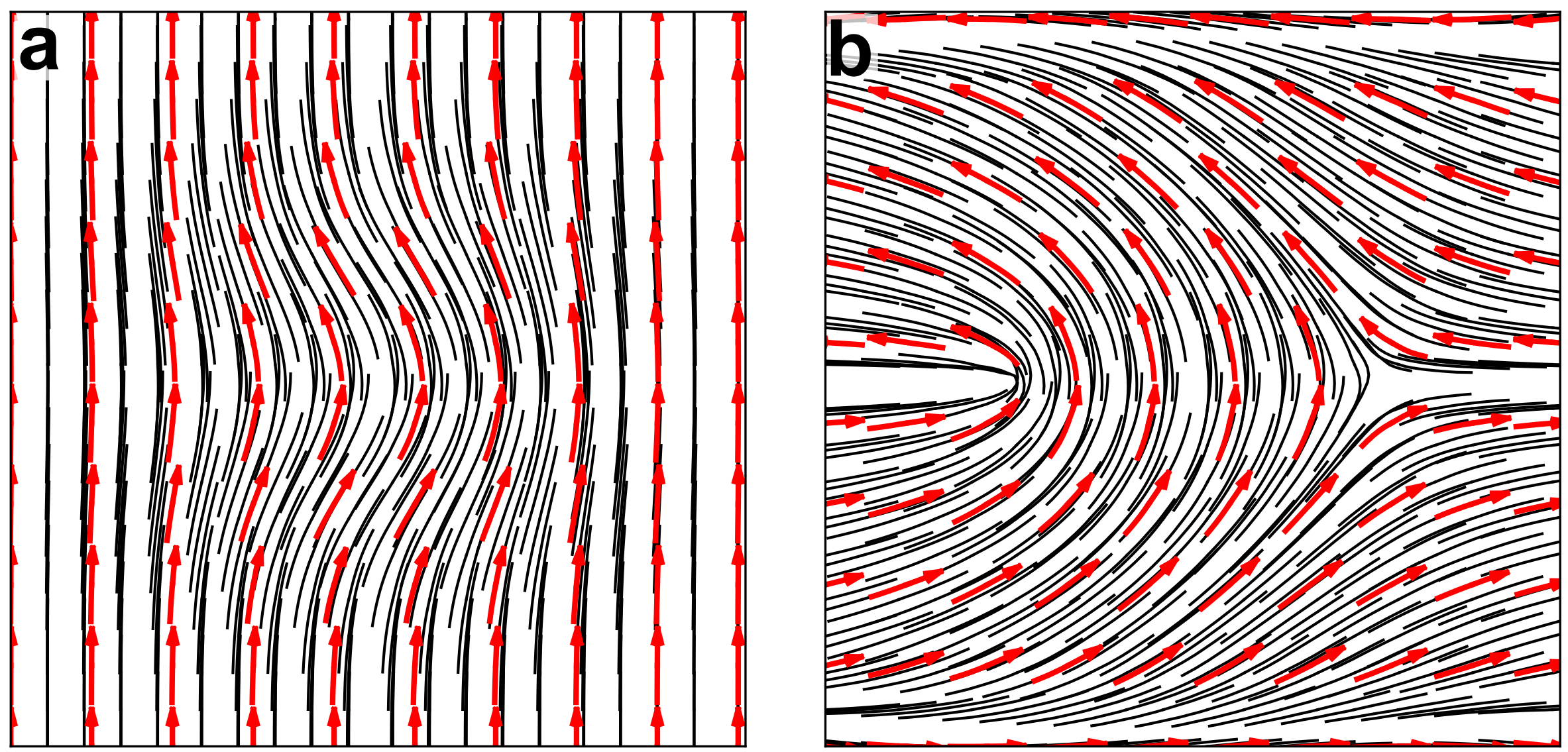}
    
    
    \captionsetup{justification=justified, singlelinecheck=false}
    \caption*{\textbf{Extended Data Fig. 7: Initial conditions for simulations.}    \red{Initial conditions for simulations of (\textbf{a}) creation, given by Eq.~\ref{eq:in_cre} and (\textbf{b}) annihilation, given by Eq.~\ref{eq:in_ann}.
    }}
    \label{fig:S7}
\end{figure*}

\begin{table*}
    \centering
    \red{
    \caption{\red{\textbf{Entropy production (EP) estimation - control.} Top: EP for trajectories of individual bacteria and cells. Bottom: EP for the nematic and flow fields at random positions. Values show the estimate $\pm$ the standard error. Since this is a lower bound, a value that is statistically larger than zero implies non reversible dynamics. See the methods section for details.}}
    \label{tab:table1} 
    \begin{tabular}{c|c|c}
        \hline
        & \textbf{Bacteria} & \textbf{Cells} \\
        \hline
        EP individual trajectories   & $-0.004\pm 0.004$ & $0.003\pm 0.0002$ \\
        EP nematic+flow & $0.0008\pm 0.001$ & $0.0008\pm0.001$ \\
        \hline
    \end{tabular}
    }
\end{table*}

\end{center}
\end{document}